\documentclass[a4paper,fleqn,usenatbib,useAMS]{mnras}
\usepackage[T1]{fontenc}
\usepackage{ae,aecompl}
\usepackage{txfonts}
\usepackage{multirow}
\usepackage{graphicx}
\usepackage{xcolor}
\usepackage{epsfig,graphics}
\usepackage{rotating} 
\usepackage{savefnmark}

\newcommand {\be}{\begin{equation}}
\newcommand {\ee}{\end{equation}}  
\renewcommand{\theenumi}{\arabic{enumi}.}

\renewcommand{\deg}{^{\circ}}
\renewcommand{\ss}{SS433}

\title[Reflection of SS 433 X-ray emission]{Is SS 433 a misaligned ultraluminous X-ray source?\\ Constraints from its reflected signal in the Galactic plane}   
\author[Khabibullin \& Sazonov]{
I. Khabibullin$^{1}$\thanks{khabibullin@iki.rssi.ru}\& 
S. Sazonov$^{1,2}$\thanks{sazonov@iki.rssi.ru}\\
$^{1}$Space Research Institute, Russian Academy of Sciences,
Profsoyuznaya 84/32, 117997 Moscow, Russia\\
$^{2}$Moscow Institute of Physics and Technology, Institutsky
per. 9, 141700 Dolgoprudny, Russia
}
\date{Received \today}
\begin{document}
\maketitle
\begin{abstract}

We evaluate the emission that must arise due to reflection of the putative collimated X-ray radiation of SS 433 by atomic gas and molecular clouds in the Galactic plane and compare the predicted signal with existing \textit{RXTE} and \textit{ASCA} data for the region of interest. Assuming that the intrinsic X-ray spectrum of SS~433 is similar to that of ultraluminous X-ray sources (ULXs), we obtain an upper limit of $\sim 2\times 10^{39}$~erg~s$^{-1}$ on its total (angular-integrated) luminosity in the 2--10 keV energy band, which is only weakly dependent on the half-opening angle, $\Theta_r$, of the emission cone. In contrast, the upper limit on the apparent luminosity of SS~433 (that would be perceived by an observer looking at its supercritical accretion disk face-on) decreases with increasing $\Theta_r$ and is $\sim 3\times 10^{40}$~erg~s$^{-1}$ for $\Theta_r\gtrsim\Theta_p=21\deg$, where $\Theta_p$ is the precession angle of the baryonic jets (assuming that the emission cones precess in the same manner as the jets). This leaves open the possibility that SS~433 is a misaligned ULX. Further investigation of the reflection signal from the molecular clouds using higher angular resolution observations could improve these constraints with the potential to break the degeneracy between $ \Theta_r $ and the apparent luminosity.
\end{abstract}
\begin{keywords}
X-rays: binaries -- X-rays: individual(SS 433)
\end{keywords}
\section{Introduction}
\label{s:intro}

~~~~~~ X-ray emission of normal galaxies consists of thermal radiation of hot ($ \sim0.5 $ keV) interstellar gas and cumulative radiation of X-ray binaries (XRBs), including low-mass and high-mass ones (LMXBs and HMXBs), the former being dominant in elliptical galaxies and the latter in spiral ones (see e.g. \citealt{Fabbiano2006} and \citealt{GilfanovGS2004} for reviews). In our Galaxy, there is also a thin, apparently diffuse X-ray emission along the Galactic plane -- the Galactic ridge X-ray emission (GRXE, \citealt{Worrall1982}), at least $ \sim 80\% $ of which is actually provided by faint point sources such as accreting white dwarfs and coronally active stars \citep{Revnivtsev2006,Revnivtsev2009}. Some fraction of GRXE may arise from scattering of radiation from XRBs by atomic gas and molecular clouds in the Galactic plane \citep{Sunyaev1993}. Recently, \cite{Molaro2014} performed a global calculation of this emission across the Milky Way based on the known properties of the XRB populations in our and other galaxies, assuming XRBs to be isotropic sources.  

{In the standard picture} of disc accretion \citep{SS1973}, it is natural to expect some degree of collimation of XRB radiation along the axis of the accretion disc (and its corona), in particular in objects accreting close to or above the Eddington limit, such as those populating the high-luminosity end of the XRB luminosity function \citep{Grimm2002,Grimm2003,Lehmer2010,Mineo2012}.  In particular, the brightest extragalactic XRBs -- ultraluminous X-ray sources (ULXs), with apparent luminosities in excess of $\sim 2\times10^{39} $ erg/s (see \citealt{Feng2011} for a review), might be super-Eddington accretors whose accretion disc is viewed face-on \citep{Fabrika2001,Rappaport2005,Begelman2006,Poutanen2007,Fabrika2015}. If such sources were observed edge-on, they might appear orders of magnitude fainter in X-rays as a result of obscuration of the luminous inner regions by a geometrically thick disc and (or) wind. {There are no known ULXs in the Galaxy, but there may be 'misaligned' ULXs present, and their beamed X-ray emission may} illuminate and scatter off the atomic gas and molecular clouds in the Galaxy and thus give rise to potentially observable 'X-ray echoes'. 

Such a situation might be relevant for the prototypical supercritical accretor and unique Galactic XRB SS 433. Its permanent accretion rate is highly supercritical, $ \dot{m}=\dot{M}/\dot{M}_{cr}\sim 400$ (where $\dot{M}_{cr}$ is the critical accretion rate), and it is viewed nearly edge-on, so that the inner parts of the supercritical accretion disc are not directly observable \citep{Fabrika2004}. Its apparent X-ray luminosity ($ \sim 10^{36} $ erg/s) is dominated by thermal emission from a pair of mildly relativistic baryonic jets \citep{Kotani1996}. Apart from this compact X-ray emission, the jets reveal themselves at radio, optical and X-ray wavelengths over a huge range of angular scales (see \citealt{Fabrika2004} for a review) and also through their impact on the surrounding SS 433 radio nebula W50 \citep{Dubner1998}. 

It is natural to expect that the funnel of the supercritical accretion disc and thus the putative collimated X-ray emission cone(s) in SS 433 are pointing in the same direction as the jets, whose time-dependent orientation is known very well. This makes it possible to estimate the reflected signal produced by this X-ray beam in the Galactic plane, depending on intrinsic properties of the collimated emission such as the time-averaged luminosity and the opening angle of the emission cone. The reflected signal should also depend on the properties of the interstellar environment of SS 433, in particular on the presence of molecular clouds in the illuminated region, and on the actual distance to SS 433, which may lie between 4.5 kpc \citep{Marshall2013} and 5.5 kpc \citep{Blundell2004}. 

By comparing our predictions with available X-ray data, we put constraints on the parameters of SS 433 collimated X-ray emission and discuss them in the context of ULXs and supercritical accretors in general.

\section{Geometry of the problem}
\label{s:geometry}

Reflection of hard X-ray (at energies above a few keV) emission by atomic and molecular interstellar gas is well understood and arises in a large variety of astrophysical situations (see \citealt{Sunyaev1996} for a related discussion). Particularly well known is the 'X-ray echo' of past X-ray activity of the Galactic supermassive black hole associated with giant molecular clouds in the Galactic Centre region \citep{Sunyaev1993,Koyama1996,RevnivtsevGC2004,Ponti2010}. In the present study, we are dealing with a somewhat more complex geometry of the illuminated region. In what follows, we assume that the axis of the SS 433 emission cone coincides with the instantaneous direction of its compact X-ray jets (as is natural to expect in the highly super-Eddingtion regime, when the radiation field likely plays the main role in launching the jets, see e.g. \citealt{Ohsuga2014}), while its half-opening angle is a free parameter of the model. 

\subsection{The source}
\label{ss:source}

The SS 433 jets precess {with amplitude $ \Theta_p=20.92\pm0.08\approx 21\deg $ and period $ P_{p}=162.375\pm0.011\approx 162$ days} around an axis inclined at angle $ i=78.05\pm0.05\approx78\deg$ to the line of sight \citep{Eikenberry2001}. Since the putative collimated X-ray radiation is never observed directly from SS 433, the half-opening angle $ \Theta_r$ of the emission cone cannot be larger than $ i-\Theta_p\approx 57\deg$. Taking into account that the jets are also subject to nutation `nodding' {with amplitude $ \Theta_n=2.8\pm0.3\approx 3\deg $} \citep{Fabrika2004}, we can further tighten this constraint: $\Theta_r<\Theta_{r,max}=i-(\Theta_p+\Theta_n)\approx 54\deg$.

Propagation of the SS~433 jets can be traced over a large range of angular {distances: through} observations of radio emission at the scale of tens of milliarcseconds (e.g. \citealt{Marshall2013}), 'corkscrew'-shaped radio and X-ray emission at arcsecond scales (e.g. \citealt{Miller2008}), and finally, at the scale of tens of arcminutes, extended X-ray emission and the `hot-spots'  (e.g. \citealt{Brinkmann1996}), which are believed to mark the termination region of the jets approximately $R_{W50}\sim100$ pc from the central source, where the bulk of the jets' kinetic energy is lost as a result of interaction with the encompassing W50 nebula \citep{Goodall2011}. 

These manifestations provide a record of SS 433 activity: assuming that the jet velocity $ V\approx0.26~c $ varies only slightly over the source's lifetime, we infer that it has been active for at least $\tau\sim R_{W50}/V\sim 1000$ years. Therefore, the potential X-ray illuminated region has a size of at least $ R_{sc}\sim\tau c\sim 400$~pc, which is more than the distance from SS 433 {to the Galactic plane, $ h\sim 175-215 $ pc (for the line-of-sight distance to SS 433 ranging from 4.5 to 5.5 kpc, see Section \ref{sss:environs})}, where the bulk of atomic and molecular gas is concentrated {(see Sections \ref{sss:atomic} and \ref{sss:molecular})}.

\begin{figure}
\includegraphics[width=0.8\columnwidth,angle=0]{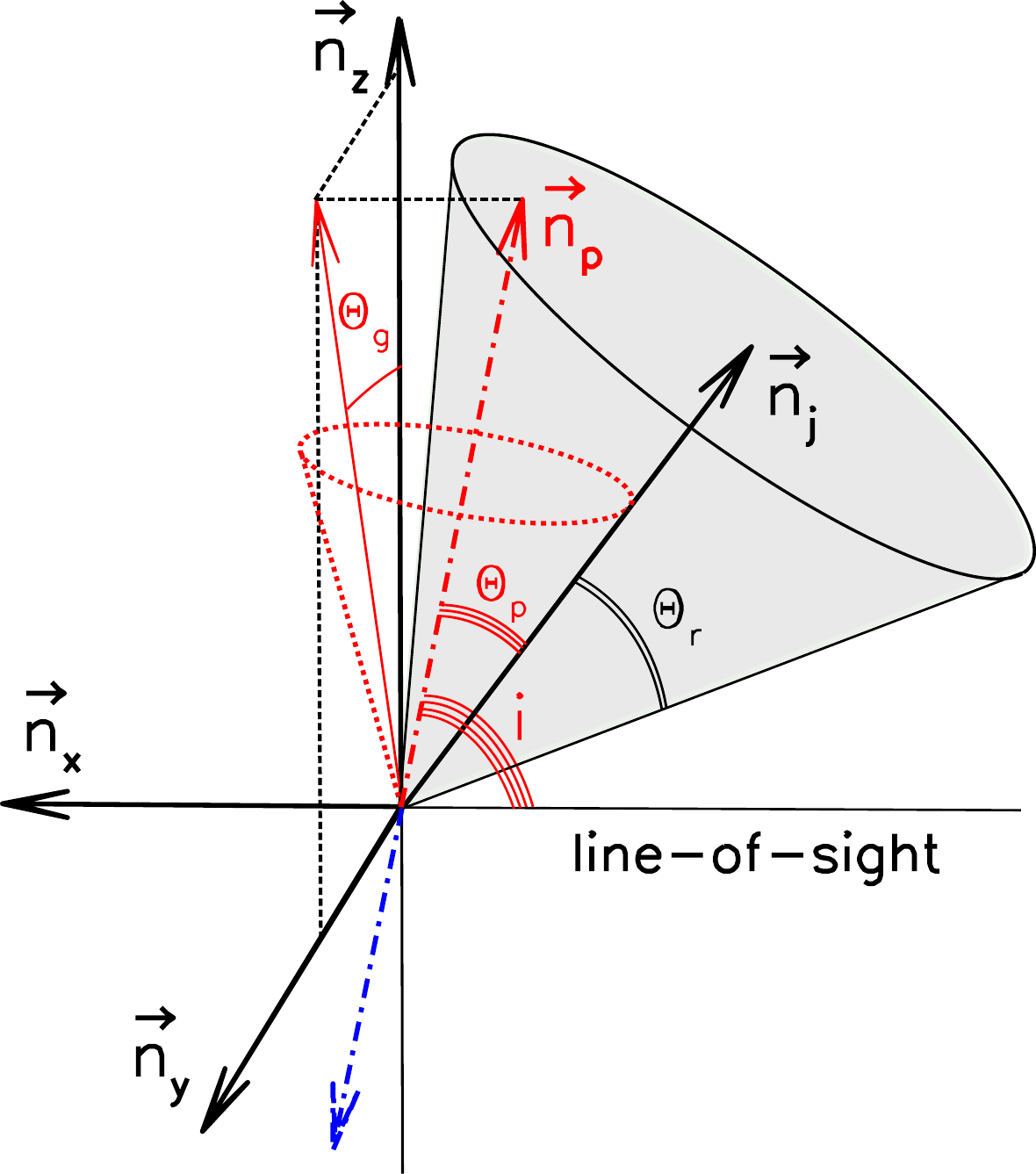}
\caption{Assumed geometry of the collimated X-ray emission from SS~433. Vector $ \vec{n_x}$ is directed towards the observer, $ \vec{n_z}$ is aligned with the positive normal to the Galactic plane and together with vector $ \vec{n_y}$ defines the projection plane (disregarding here the small angular offset of SS~433 from the Galactic plane). The direction of the precession axis is marked with red ($ \vec{n_p}$, for the receding jet) and blue (for the approaching jet) dash-dotted arrows. The projection of $ \vec{n_p}$ on the ($ \vec{n_y}$,$ \vec{n_z}$) plane makes an angle $ \Theta_g \approx20\deg$ to $ \vec{n_z} $, while the angle between  $\vec{n_p}$ and $ \vec{n_x}$ equals $ \pi-\rm{i}\approx 102\deg $. The instant direction of the receding jet is shown by the $ \vec{n_j}$ vector making an angle $ \Theta_p\approx21\deg$ to $ \vec{n_p}$, {and the red dotted cone depicts the precession cone of the jet's direction. The instant region of illumination is shown by the grey-shaded cone with half-opening angle $ \Theta_r$, and it evolves in time accordingly, i.e. it also precesses with period $P_p\approx162  $ days.}
} 
\label{f:angles}
\end{figure}

The 3D direction of the jets can be fully reconstructed with no ambiguity between the (mostly) approaching and (mostly) receding jet. Indeed, the projection of the precession axis on the sky is inclined at {an angle $ \Theta_g\approx20\deg $} (towards the Galactic centre) relative to the normal to the Galactic plane \citep{Goodall2011}. The full angle between the normal to the Galactic plane and the jet precession axis {is thus $\Theta_{GP}={\rm cos^{-1}}(\sin i~\cos\Theta_{g})\approx 23\deg$}. It is the \textit{receding} jet that propagates towards the Galactic plane, as inferred from fitting the kinematic model to the corkscrew-shaped pattern of radio emission at arcsecond scales (e.g. \citealt{Miller2008}) and confirmed (though with less confidence) by the redshifts of spectral lines appearing in the spectrum of the extended X-ray emission observed by \textit{Chandra} at the same scales \citep{Migliari2002}\footnote{In the paper by \cite{Migliari2002}, the 7.06 keV line of Fe I K$ \beta$ is erroneously referred to as the Fe XXV K$ \beta$ (7.89 keV) line.}. The geometry of the emission cone derived from these considerations is presented in Fig.~\ref{f:angles}.


\subsection{The illuminated region}
\label{ss:illuminationregion}

Due to precession of the X-ray emission cone, the geometry of the illuminated region is rather complicated and time-dependent, but periodical in time and (approximately) in space (along the line connecting the source with the illuminated points) { with the periods $ P_p $ and $ z_p=cP_p\approx0.136$ pc (for distances from the source much larger than $z_p$), respectively. Since $ z_p $ is much smaller} than the characteristic scale-height of the atomic gas distribution in the Galaxy ($ z_d\sim 40$ pc, see {Section \ref{sss:atomic}}) and typical sizes of molecular clouds ($ r_{m}\sim 1$--20~pc, \citealt{RD2010}), the reflected emission from such structures should be sensitive only to the fraction of time they are exposed to X-rays from SS~433 and its luminosity averaged over the corresponding light-crossing time ($z_d/c\sim 130$ yrs or $ r_m/c\sim 3$--60~yrs).

\subsubsection{Illumination by precessing collimated emission}
\label{sss:precession}

\begin{figure}
\includegraphics[bb=40 180 600 670, width=1.05\columnwidth,angle=0]{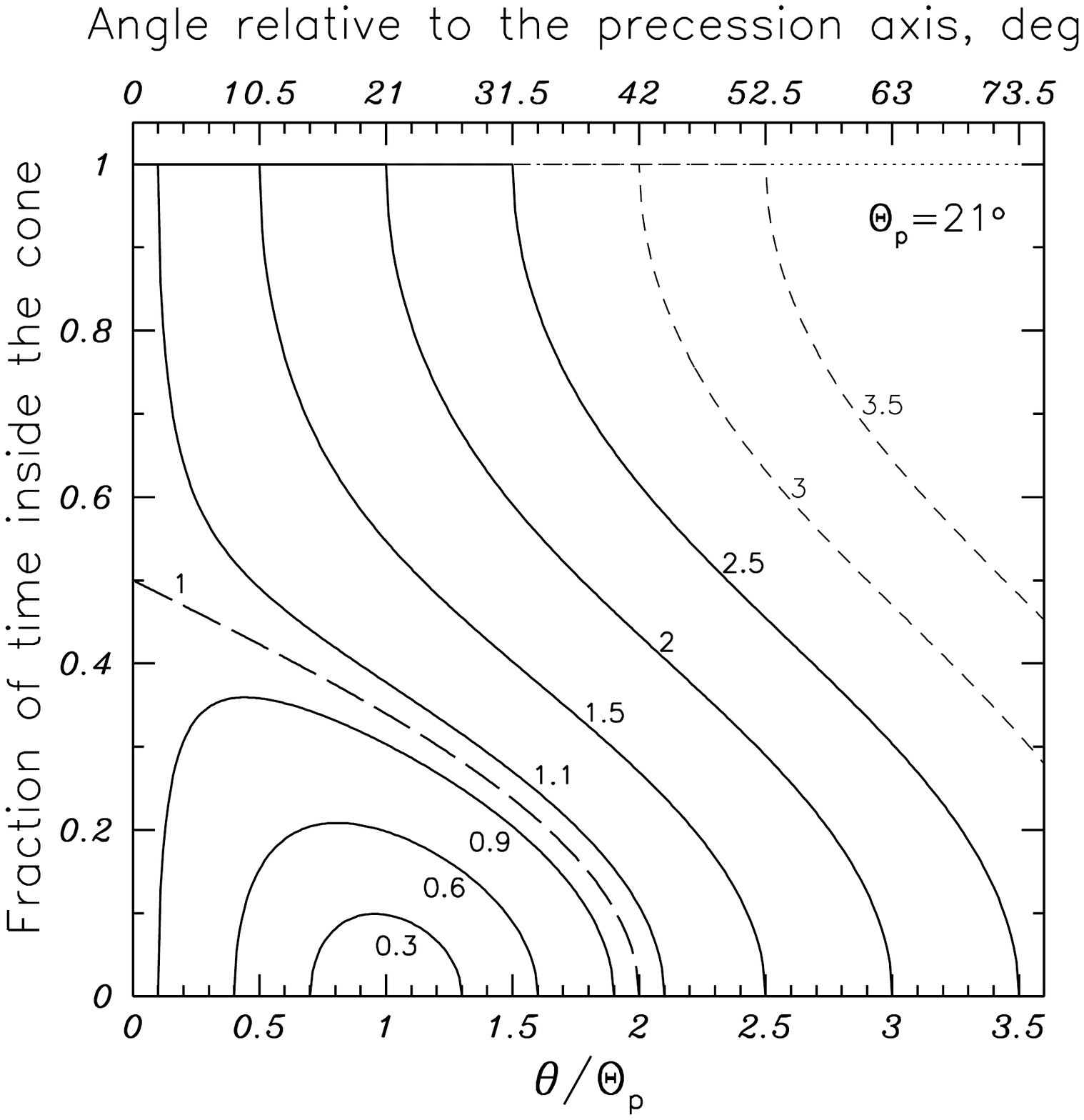}
\includegraphics[bb=40 400 600 690, width=1.05\columnwidth,angle=0]{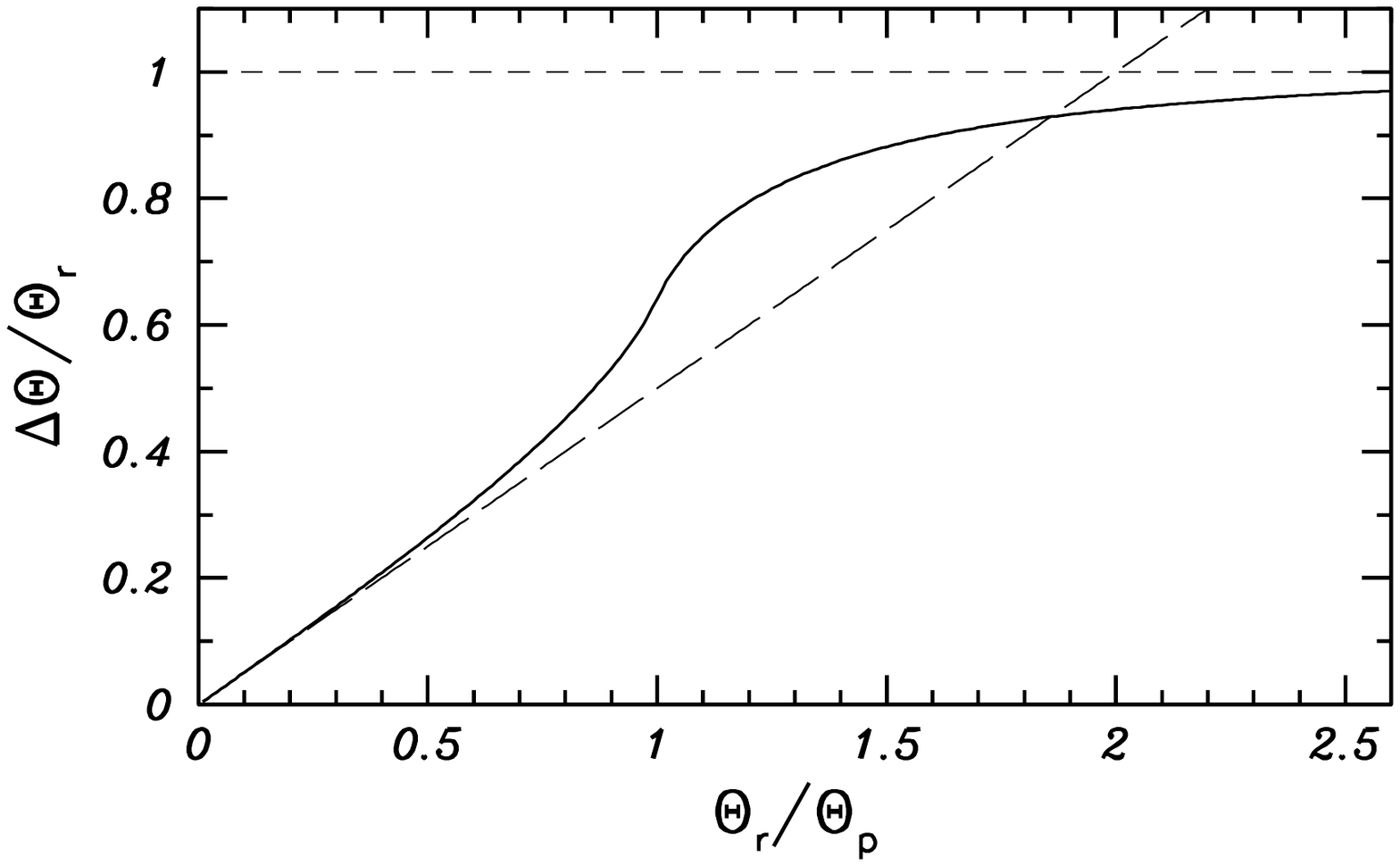}
\caption{\textit{Top panel.} Fraction of time, $F(\theta)$, a given direction (defined by the angle $ \theta $ it makes with the precession axis) spends inside the precessing (with amplitude $\Theta_p=21\deg$) collimated emission cone with half-opening angle $ \Theta_r$. Various curves correspond to various $ \Theta_r$ as indicated by values of $ \eta=\Theta_r/\Theta_p$ next to the curves. The long-dashed curve ($ \eta=1$) marks the separatrix of two regimes of illumination (see text), while the short-dashed curves (for $ \eta=3$ and $ 3.5$) correspond to values of $ \Theta_r $ that do not satisfy the $ \Theta_r\leqslant i-(\Theta_p+\Theta_n)\approx 54\deg$ condition and are thus ruled out for SS~433. \textit{Bottom panel.} Effective angular width $ \Delta\Theta $ (in units of $ \Theta_r $) found by integration of $ F(\theta)$ over all $ \theta$ as a function of $ \eta $ (solid curve). It approaches the asymptotes $\Delta\Theta/\Theta_r=0.5\eta$ (the long-dashed line) at $\eta\ll 1$ and $\Delta\Theta/\Theta_r=1$ (the short-dashed line) at $\eta\gg 1$. }
\label{f:frac}
\end{figure}

Due to the axial symmetry with respect to the precession axis, the
duty cycle, $F$, of illumination of a given position is a function of
the angle $ \theta $ between the line connecting this position with
the source and the precession axis (see Fig.~\ref{f:frac}). {
  It is readily calculated as a fraction of time a given point on a
  sphere spends inside a circle with angular radius $\Theta_r $ that
  rotates around an axis at angular distance $\Theta_p $ from the
  circle's center. Naturally, the shape of the $F(\theta)$ function
  depends on $\Theta_p $ and $\Theta_r $: }
\begin{center}
\begin{tabular}{c}
$ F(\theta)=\frac{1}{\pi}\cos^{-1}\left[\xi_{\Theta_p,\Theta_r}(\theta)\right]$, ~~ if~~ $ |\xi_{\Theta_p,\Theta_r}(\theta)|\leq 1$,
\end{tabular}
\end{center}
\begin{equation}
 F(\theta)=1,~~~ \textrm{if}~~  \xi_{\Theta_p,\Theta_r}(\theta)\leq -1 ~\textrm{and}~
\end{equation}
\begin{center}
\begin{tabular}{c}
$ F(\theta)=0$, ~~ if~~  $ \xi_{\Theta_p,\Theta_r}(\theta)\geq 1$, 
\end{tabular}
\end{center}
where
\begin{equation}
\xi_{\Theta_p,\Theta_r}(\theta)=\frac{\cos\Theta_r-\cos\theta\cos\Theta_p}{\sin\theta\sin\Theta_p}.
\end{equation}
{For small $\Theta_p $, $\Theta_r $ and $ \theta $, the shape of $ \xi_{\Theta_p,\Theta_r} $ is determined only by the ratio $ \eta=\Theta_r/\Theta_p $:}
\begin{equation}
\xi_{\Theta_p,\Theta_r}(\theta)\approx\xi_{\eta}(\theta)=\frac{1-\eta^2 +(\theta/\Theta_{p})^2}{2(\theta/\Theta_p)}.
\end{equation}

One may therefore distinguish two regimes of illumination. If the
emission from SS~433 is weakly collimated ($ \Theta_r>\Theta_p$,
i.e. $ \eta>1 $), {every location within a cone with the half-opening angle $\Theta_1=\Theta_r-\Theta_p$ and its axis coinciding with the precession axis} is \textit{always} illuminated by SS~433 despite the precession. Outside this cone, the fraction of `on time' decreases rapidly, {reduced by half} at an angle $ \Theta_{0.5}\sim \Theta_1+\Theta_p/2$ from the precession axis (see Fig.~\ref{f:frac}). In the opposite regime of strongly collimated emission ($ \Theta_r<\Theta_p$), the fraction of {the illuminated} time does not exceed 1/2 for any position, and the region of illumination is a hollow cone with the opening angle $ \approx\Theta_p$ and angular thickness of the `walls' $ 2\Theta_r$. For small $ \eta=\Theta_r/\Theta_p $, $F(\theta) $ reaches a maximum of $\approx \eta/\pi$ {at $ \theta\approx\Theta_p $ (see Fig. \ref{f:frac})}.

It is convenient to introduce an effective angular {half-width} 
\begin{equation}
\Delta\Theta=\int_{0}^{\pi/2}F(\theta)d\theta,
\label{eq:effangle}
\end{equation}
which translates into an effective {half-width} of the scattering zone along the line of sight when dealing with an infinite homogeneous medium and predicting the surface brightness of the reflected signal (see eq.~(\ref{eq:dd}) below). In the wide-beam limit ($\Theta_r\gg\Theta_p$), $\Delta\Theta\approx\Theta_r$, and in the narrow-beam limit ($\Theta_r\ll\Theta_p$), $\Delta\Theta\approx \Theta_r^2/2\Theta_p $ (see Fig.~\ref{f:frac}). 

If the scatterer covers {half of the whole sky when 'viewed' from SS~433 and if the scattering opacity is constant over the photons energy range of interest (as is approximately the case for the atomic gas in the Galactic plane and X-rays with energy >4 keV, see Section \ref{s:prediction}), then the total scattered luminosity $L_{sc}$  does not depend on the precession pattern}, and is determined solely by its total, i.e. angular-integrated, luminosity, $L_c$, and the optical depth of the scatterer, $ \tau_{sc} $, with respect to photons emitted by the central source:
\begin{equation}
{L_{sc}}=\frac{L_c\tau_{sc}}{2}=\frac{\Omega_{r}}{4\pi}L_0\tau_{sc},
\end{equation}
where $\Omega_r=2\pi(1-\cos\Theta_r)$ is the solid angle of the emission cone (actually assuming that there are two opposite identical emission cones) and $L_0$ is the 'apparent', or isotropic equivalent, luminosity (as would be estimated by an observer looking into the SS~433 emission cone by multiplying the measured X-ray flux by $4\pi d^2$, {with $ d $ being the assumed distance to the source}). 

\subsubsection{Location of the illuminated region {within the Galactic plane}}
\label{sss:environs}
\begin{figure*}
\includegraphics[width=0.9\columnwidth,angle=0]{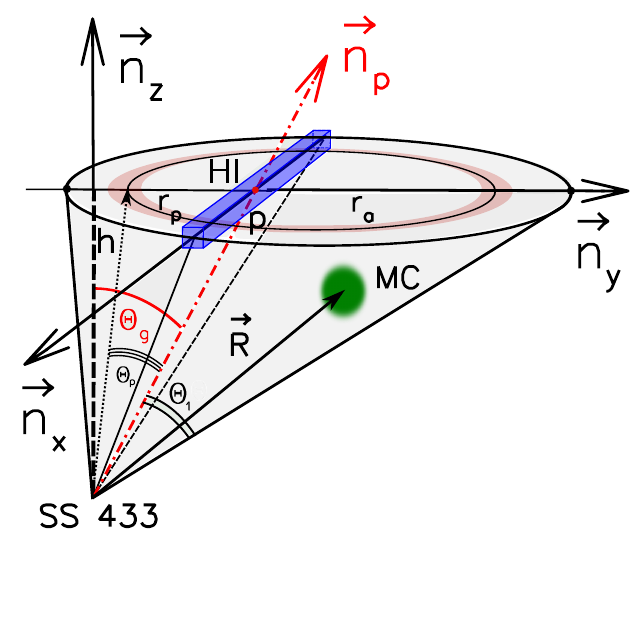}
\includegraphics[bb=0 170 560 710,width=1.1\columnwidth,angle=0]{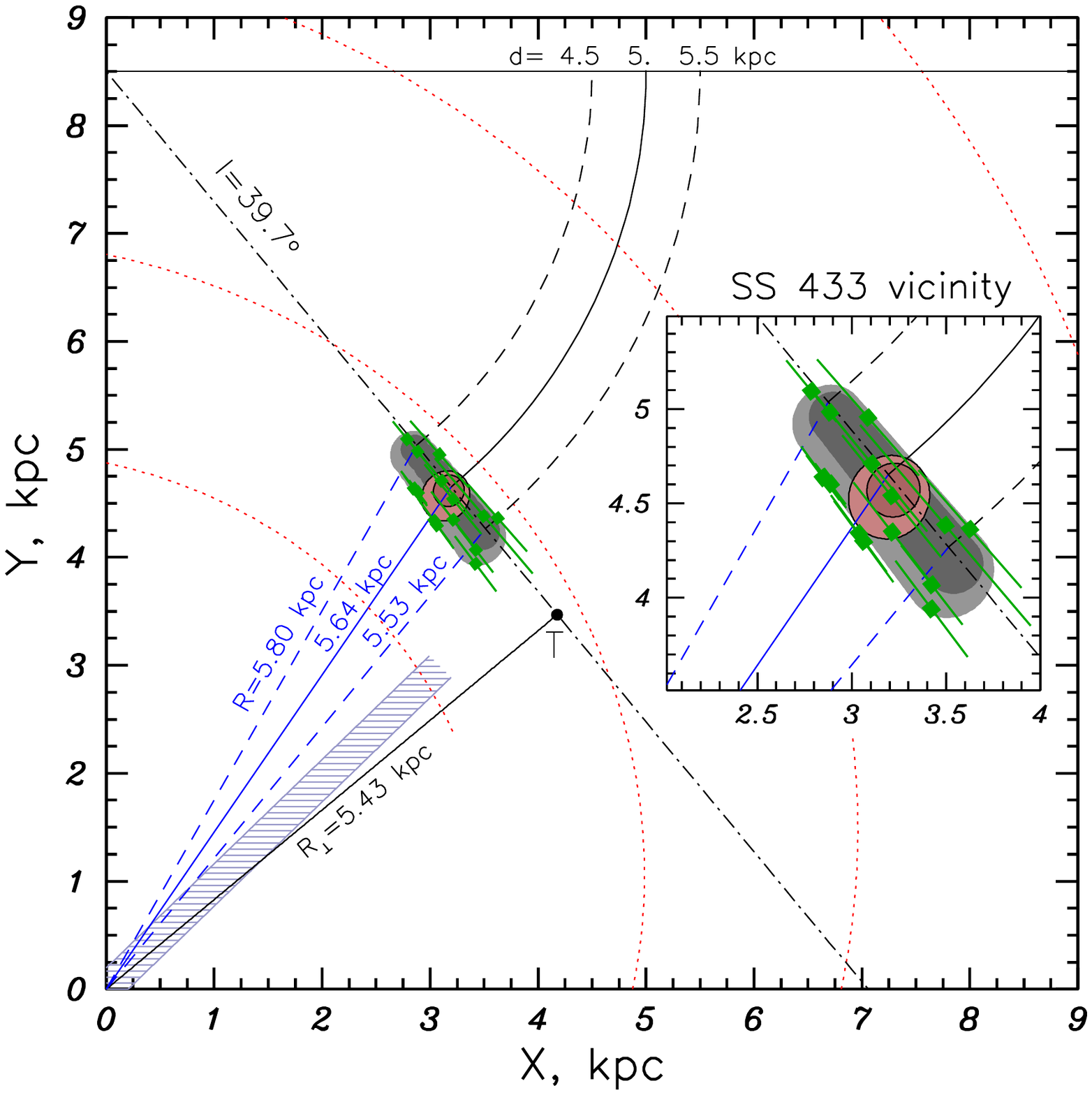}
\caption{\textit{Left panel.} Schematic of the illumination of atomic
  interstellar gas (labelled 'HI') and {a molecular cloud}
  (labelled 'MC') by the collimated X-ray emission from SS 433
  {(the inclination angle \textit{i} of the precession axis to
    the line of sight is taken equal to $ 90\deg$ for
    simplicity)}. The gray cone shows the region of permanent
  illumination ($ \theta<\Theta_{1}$) in the case of a wide emission
  cone ($\Theta_r\lesssim 2\Theta_p$). {In the case of a narrow
    emission cone ($ \Theta_r\ll\Theta_p $), the illuminated (for only
    some fraction of time) region is essentially a hollow cone (see
    text), whose intersection with the Galactic plane is shown by the
    reddish ring.} 
The directions of the vectors $ \vec{n_x}$, $ \vec{n_y}$, $ \vec{n_z}$ and $ \vec{n_p}$ and designations of the angles are the same as for Fig.~\ref{f:angles}. \textit{Right panel.} Projection of the positions of SS~433 and potentially illuminated molecular clouds on the Galactic plane. The Galactic Centre (GC) is located at the coordinate origin, while the observer at (0, 8.5 kpc). The direction from the observer to SS 433 (the $ l=39.7\deg $ line) is shown as the dash-dotted line,{ and the location marked 'T' on this line is the tangent point of the Galactic rotation curve for this direction}. The arcs centred on the observer's position mark the allowed SS~433 distance range from $ d_{\ss}=$4.5 to 5.5 kpc (as indicated next to their upper endpoints). The corresponding Galactocentric distances are indicated next to the blue lines originating from the GC. Approximate locations of the spiral arms (according to the prescription by \citealt{Vallee1995}) are shown by red dotted lines, while the location of the `long' Galactic bar is indicated by the gray dashed bar extending mostly up to $ l=30\deg $ \citep{Benjamin2005}. The small light-shaded ellipse shows, for $ d_{\ss}=5 $ kpc, the region that will be constantly illuminated by SS~433 in the case of the largest possible half-opening angle of the emission cone of $ \Theta_r=54\deg $ (so that $ \Theta_{1,max}=33\deg$). The small elongated shaded region corresponds to the same $ \Theta_r$ but indicates the uncertainty in the position of the illuminated region due to the uncertainty in $ d_{\ss} $: from 4.5 to 5.5 kpc. The larger shaded ellipse and elongated region show (approximately) the corresponding regions of illumination during at least 50\% of time. Green dots with the sticks show the positions of the relevant molecular clouds and corresponding uncertainties in distances to them (see Table~\ref{t:mcs}). }
\label{f:lands}
\end{figure*}

SS 433 has Galactic coordinates $l_{\ss}=39.6940{\deg}\approx 39.7\deg$, $ b_{\ss}=-2.2445{\deg}\approx-2.2\deg$ {and is located approximately $ d_{\ss}\sim 5$~kpc from {the Sun} (e.g. \citealt{Fabrika2004}). Hence, the distance from SS 433 to the Galactic plane is $ h=d_{\ss}|\sin b_{\ss}| \approx 200 d_5$ pc (where $ d_5=d_{\ss}/5$~kpc)}, while the distance to the Galactic plane along the direction of the precession axis {is slightly larger}: $ R_0=h/\cos\Theta_{GP}\approx 210~ d_5~$ pc. Figure~\ref{f:lands} depicts the orientation of the emission cone and the location and approximate shape of the illuminated region with respect to the Galactic coordinate system for both regimes described above.  

In the case of a wide emission beam {($\Theta_r>\Theta_p$)},
the section of the constantly illuminated region ($ \theta<\Theta_1 $)
by the Galactic plane has the shape of an ellipse elongated
approximately in the direction of the Galactic Centre and having the
intersection point of the precession axis with the Galactic plane
{($ l\approx38.9 \deg $, $b=0 \deg $)} in one of its foci (see
Fig.~\ref{f:lands}). The ellipse's major axis equals $ 2a=r_a+r_p $,
with $r_a= h\left[\tan(\Theta_1+\Theta_{GP})-\tan \Theta_{GP})\right]$
(apocenter distance) and $r_p=h\left[\tan(\Theta_1-\Theta_{GP})+\tan
  \Theta_{GP}\right] $ (pericenter distance), {and since $
  \frac{\pi}{2}-i \approx 12\deg$ is rather small, this axis is roughly perpendicular to the line of sight.} For the largest possible half-opening angle of the SS~433 emission {cone of $ \Theta_{r,max}=54\deg $, $\Theta_{1,max}=\Theta_{r,max}-\Theta_p=33\deg $}, $ r_a\approx 1.06~ h $ and $ r_p\approx 0.60 ~h $, so $ 2a\approx 1.7~h\approx 340~ d_5 $ pc. { Hence, the constantly illuminated region has a maximal angular width $ \Delta l\approx2a~\sin i~/d_{\ss}\approx 3.7$ deg} in the Galactic plane. Similarly, for the region illuminated at least 50\% of time ($\theta<\Theta_{0.5}$), $\Theta_{0.5,max}=\Theta_{r,max}-\Theta_p/2.=43.5\deg $, $ r_a\approx 1.9~ h $ and $ r_p\approx 0.8 ~h $, so $ 2a\approx 2.7~h\approx 540~ d_5 $ pc and $ \Delta l\approx2a~\sin i~/d\approx 6$ deg. The 'width' (i.e. the extent in the direction perpendicular to the major axis and approximately along the line of sight) of the ellipse at the focus {is $ 2p=\frac{4r_ar_p}{r_a+r_p}\approx 2R_0\tan\Theta_1\approx 280 d_5$ pc} for $\Theta_1=\Theta_{1,max}$ and $ \approx 412 d_5 $ pc for $ \Theta_{0.5}= \Theta_{0.5,max}$. The largest 'width' of the ellipse is $\Delta d=2\sqrt{r_ar_p} $ and is shifted {from the main focus by $(r_a-r_p)/2=a e\ll a$}, since the ellipse's eccentricity $ e $ is rather small for the relevant values of $ \Theta_1 $ and $ \Theta_{GP} $. 

If the reflecting medium is homogeneous inside the illuminated region, the scattering column density is effectively collected within
\begin{equation}
 \Delta d\approx 2 R_0 \tan\Delta\Theta\approx 170 \frac{\tan \Delta\Theta}{\tan \Theta_p}~{\rm pc}
 \label{eq:dd}
\end{equation}
along the line of sight, where $\Delta\Theta$ was defined in equation~(\ref{eq:effangle}) and shown in Fig.~\ref{f:frac}. 

In the case of a narrow emission beam ($\Theta_r<\Theta_p$), the illuminated region in the Galactic plane is a ring of a similar ellipse, but now determined by the section of the precession cone by the Galactic plane, with characteristic 'width' also given by equation~(\ref{eq:dd}) with $\tan \Delta\Theta/\tan \Theta_p\approx 0.5\eta^2 $  (see Fig.~\ref{f:lands}). In this case, the projection of the illuminated region is approximately confined by the points of intersection of the precession cone with the Galactic plane {located at $(l\approx 37.9 \deg,~b=0\deg) $ and $( l\approx 39.8 \deg,~b=0\deg) $} (see also Fig.~\ref{f:xtemap}). 

\subsubsection{Atomic gas}
\label{sss:atomic}

The distribution of atomic interstellar gas in the Galactic disc is rather smooth and well described by an exponential disc model 
with a vertical scale-height $ z_d=40 $ pc \citep{Dehnen1998}, normalized so that the $n_{HI} $ density in the Solar vicinity is $\approx 0.9$~cm$ ^{-3}$ \citep{Kalberla2009}. Still, there are large-scale perturbations in the distribution of the atomic gas, first of all related to the spiral arms and the long Galactic bar. However, as shown in Fig.~\ref{f:lands} (right panel), these structures are well beyond the potential illumination region in our problem, so these deviations from the smooth distribution can be neglected. The Galactocentric distance of SS 433 varies from 5.80 to 5.53 kpc for $ d_{\ss} $ in the range from 4.5 to 5.5 kpc {(see Fig.~\ref{f:lands})}, and the $ n_{HI}$ density in the illuminated part of the Galactic disc is $ \sim 1 $ cm$ ^{-3}$ (cf. \citealt{Goodall2011}). There are also small-scale structures in the HI distribution, but since the SS~433-illuminated region is fairly large ($\sim 100 $ pc), these density inhomogeneities may also be considered unimportant for the problem at hand. 

\subsubsection{Molecular clouds}
\label{sss:molecular}

Contrary to the atomic gas, the molecular ISM phase is essentially clumpy, so that a significant fraction of the mass resides in dense clouds with typical overdensity of 100--1000, characteristic sizes of $ \sim10$--100~pc and a small volume filling factor ($ \sim10^{-3} $, \citealt{McKee2007}). To predict the corresponding reflected X-ray signal, we should use the actual positions and parameters of molecular clouds (MCs) in the region if interest. To this end, we use the data from the Boston University-Five College Radio Astronomy Observatory (BU-FCRAO) Galactic Ring Survey \citep{Jackson2006}. It provides a 3-dimensional (celestial position and line-of-sight velocity) map of the molecular gas with angular resolution of $ 46'' $  for $ 18\deg\leq l\leq 55.7\deg $ and $ -1\deg\leq b\leq 1\deg $, including the region of our interest. Since this survey exploits in most cases the optically thin tracer $^{13}$CO $J = 1\rightarrow 0$ rather than the optically thick $^{12}$CO $J = 1\rightarrow 0$, it is efficient at detecting and determining characteristics of the largest and densest clouds and their cores \citep{RD2010}. We use a catalogue of MCs \citep{RD2009} for which the well-known near/far ambiguity in kinematic distance estimates (based on line-of-sight velocities) was resolved using the HI self-absorption (HISA) technique. This method is based on the fact that cold HI embedded in a MC located at the near kinematic distance (i.e. closer to the observer than the tangent point of Galactic rotation) absorbs the 21 cm radiation emitted by the warm HI background located beyond the cloud, so that an HI 21 cm absorption line is observed from the cloud at the same radial velocity as the $ ^{13} $CO emission line from the cloud. As is clear from the previous discussion, the SS~433 illumination region lies entirely in the $ d<d_{tan,SS433}=d_{GC}\cos l_{\ss}\approx 6.5 $ kpc domain, where $d_{tan}$ is the distance to the tangent point in the direction of SS~433 {(i.e. for the Galactic longitude $l_{\ss}\approx39.7\deg$), and $ d_{GC}\approx8.5$ kpc is the distance to the Galactic center (see also Fig.~\ref{f:lands}, right panel)}. Therefore, all potentially interesting MCs are located at near kinematic distances. 

We selected all molecular clouds with $ 36\deg\leq l\leq 42\deg $ and distance estimates ranging from 4.3 to 5.7 kpc ({corresponding to the line-of-sight velocities ranging from} 60 to 70 km~s$^{-1}$ for $ l=36\deg $ and from 65 to 85 km~s$^{-1}$ for $ l=42\deg$), since the illuminated region may span up to $\sim 400$ pc along the line of sight and $ 6\deg $ along the Galactic plane (for $ \Theta_{0.5}=\Theta_{0.5,max}=43.5\deg $). We found 15 such clouds. Their physical parameters \citep{RD2010} are summarised in Table \ref{t:mcs}. To facilitate the presentation of our results, we divided the whole sample of MCs in two almost equal groups: 'nearby' ones, located at $ d<5 $ kpc, and 'distant' ones, located at $ d>5 $ kpc. 

\begin{table*}
\caption{Position and physical properties of MCs that may be illuminated by the collimated X-ray emission from SS 433, {arranged by distance $ d_{MC} $ to the observer (given in Column 9) and separated into two groups: 'nearby' ones with $ d_{MC}<5 $ kpc, and 'distant' ones with $ d_{MC}>5 $ kpc}. Column 2 gives the name of a cloud in the BU-FCRAO GRS catalogue \citep{RD2010}. Columns 3 and 5 give the Galactic longitude and latitude of the cloud's centroid, while columns 4 and 6 give the corresponding semi-axes of its projected sky image. Columns 7 and 8 give the radial velocity and half width at half maximum of the velocity profile. Column 10 gives $ \Delta d_{MC} $, the distance uncertainty calculated from the velocity uncertainty according to the kinematic distance formula (see text). Marked with a $ ^\bigstar$ symbol are those clouds that prove to be close in projection on the sky to some SNR (see text).}
\begin{tabular}{c|c|c|c|c|c|c|c|c|c|}
\hline
\hline
  \multicolumn{1}{|c|}{} &
  \multicolumn{1}{|c|}{GRS } &
  \multicolumn{1}{c|}{l} &
  \multicolumn{1}{c|}{$ \delta $l} &
  \multicolumn{1}{c|}{b} &
  \multicolumn{1}{c|}{$ \delta $b} &
  \multicolumn{1}{c|}{$ V_{LSR} $} &
  \multicolumn{1}{c|}{$ \delta V_{LSR} $} &
  \multicolumn{1}{c|}{$d _{MC} $} &
  \multicolumn{1}{c|}{$ \Delta d_{MC} $}\\
  \multicolumn{1}{|c|}{} &
  \multicolumn{1}{|c|}{name} &
  \multicolumn{1}{c|}{deg} &
  \multicolumn{1}{c|}{deg} &
  \multicolumn{1}{c|}{deg} &
  \multicolumn{1}{c|}{deg} &
  \multicolumn{1}{c|}{km/s} &
  \multicolumn{1}{c|}{km/s} &  
  \multicolumn{1}{c|}{kpc} &
  \multicolumn{1}{c|}{kpc}  \\  
\hline 
1&  G039.29-00.61 & 39.29 & 0.27 & -0.61 & 0.09 & 64.5 & 3.5 & 4.4 & 0.2\\
2&  $^\bigstar $G039.34-00.31 & 39.34 & 0.36 & -0.31 & 0.12 & 65.8 & 2.9 & 4.55 & 0.2\\
3&  $^\bigstar $G041.04-00.26 & 41.04 & 0.22 & -0.26 & 0.18 & 65.8 & 5.2 & 4.7 & 0.4\\
4&  G036.44+00.64 & 36.44 & 0.12 & 0.64 & 0.065 & 71.8 & 3.5 & 4.8 & 0.2\\
5&  G036.39+00.84 & 36.39 & 0.20 & 0.84 & 0.095 & 71.3 & 2.6 & 4.8 & 0.1\\
6&  G036.54+00.34 & 36.54 & 0.12 & 0.34 & 0.20 & 71.8 & 2.3 & 4.85 & 0.1\\
  \smallskip
7&  $^\bigstar $G039.34-00.26 & 39.34 & 0.29 & -0.26 & 0.49 & 69.7 & 6.2 & 4.9 & 0.5\\
8&  $^\bigstar $G039.04-00.91 & 39.04 & 0.30 & -0.91 & 0.28 & 71.8 & 5.2 & 5.1 & 0.4\\
9&  G036.14+00.09 & 36.14 & 0.10 & 0.09 & 0.08 & 75.6 & 3.9 & 5.15 & 0.2\\
10&  G036.09+00.64 & 36.09 & 0.15 & 0.64 & 0.15 & 76.5 & 4.1 & 5.2 & 0.3\\
11&  G037.74-00.46 & 37.74 & 0.44 & -0.46 & 0.21 & 74.8 & 4.3 & 5.25 & 0.3\\
12&  $^\bigstar $G040.34-00.26 & 40.34 & 0.23 & -0.26 & 0.20 & 72.2 & 5.7 & 5.4 & 0.7\\
13&  $^\bigstar $G041.24+00.39 & 41.24 & 0.31 & 0.39 & 0.30 & 71.3 & 2.2 & 5.5 & 0.4\\
14&  G037.69-00.86 & 37.69 & 0.31 & -0.86 & 0.20 & 77.7 & 2.8 & 5.6 & 0.3\\
15&  $^\bigstar $G036.89-00.41 & 36.89 & 0.21 & -0.41 & 0.18 & 79.8 & 3.5 & 5.7 & 0.3\\
\hline\end{tabular}

\label{t:mcs}
\end{table*}

Most physical parameters of MCs are not measured directly, but derived by means of certain relations between them and observables (e.g. brightness temperature in the $^{13}$CO $J = 1\rightarrow 0$ line and line-of-sight velocity). Since these relations are established by averaging over large samples of MCs under simplified assumptions, the uncertainties in derived parameter values are usually dominated by systematic effects, with some interdependence between the parameters (see \citealt{RD2010} and references therein). For instance, the {'radius' $ r $} of a MC is determined by fitting the observed shape of the cloud by an ellipse and equating the measured projected area to $ \pi (r/d)^2$, where $ d $ is the kinematic distance estimate for the cloud {(see the next paragraph}). The integrated number density of $^{13}$CO is calculated from the optical depth in the optically thin $^{13}$CO $J = 1\rightarrow 0$ line, while the total mass $ M $ of the cloud is found using the conversion relations from $^{13}$CO to $^{12}$CO and then to $ n_{H_2} $ number densities: $ n(^{12}$CO)/$n(^{13}CO)=45 $ and $ n(^{12}CO)/n_{H_2} = 8\times 10^{-5}$ (see \citealt{RD2010} and references therein). Mean gas surface densities are calculated assuming a spherical cloud with radius $ r $ and mass $M$.    

Kinematic distance estimation relies on a particular shape of the Galactic rotation curve $ V(R) $, which is close to flat ($ V(R)=V_0=220$ km/s, e.g. \citealt{Kalberla2009}) at the Galactocentric radii anticipated for SS 433 ($ R_{GC}=5.5$--5.8~kpc, see Fig.~\ref{f:lands}). For a MC located at {the Galactic longitude} $ l_{MC} $ and having line-of-sight velocity $ V_{LSR} $, the kinematic distance estimate reads
\begin{equation}
d(V_{LSR},l_{MC})=d_{tan}(l_{MC})-\sqrt{R_{GC}(V_{LSR},l_{MC})^2-d_{GC}^2 \sin^2 l_{MC}}, \label{eq:vlsr}
\end{equation} 
where $d_{GC}\simeq 8.5 $ kpc is the distance from the Sun to the Galactic centre, $ d_{tan,MC}=d_{GC}\cos l_{MC}\approx 6.5 $ kpc is the distance to the tangent point {in this direction} and
\begin{equation}
 R_{GC}(V_{LSR},l_{MC})=d_{GC} \sin l_{MC} \times\frac{V(R_{GC})}{V_{LSR}+V_{\odot} \sin l_{MC}}, 
\end{equation}
where $ V_{\odot}=V_0\approx 220 $ km/s is the radial velocity of the Sun in the Galaxy for which also a correction of 7 km/s needs to be applied to account for the peculiar motion of the Sun \citep{RD2009}. The minus sign in front of the second term in equation~(\ref{eq:vlsr}) takes into account that the MCs of interest must be located at the near (rather than far) kinematic distance. Given a FWHM of the MC radial velocity profile $\Delta V_{LSR} $, one may estimate the uncertainty in the MC distance as 
\begin{equation}
\Delta d_{MC}=\frac{1}{2}\Delta V_{LSR}\left(\frac{\partial}{\partial V_{LSR}} d(V_{LSR},l_{MC})\right)_{V=V_{LSR}},
\end{equation}
where the derivative increases rapidly as $ V_{LSR} $ approaches the tangent value for a given $ l_{MC} $ \citep{RD2009}. The so derived $\Delta d_{MC} $ values are given in Table \ref{t:mcs} and range from $\approx 130 $ to $\approx 700 $ pc.

The above estimate accounts only for the uncertainty in the MC distance caused by the way how the searching algorithm finds and describes clumps in the molecular gas distribution. An additional contribution to the distance uncertainty comes from the cloud-to-cloud velocity dispersion relative to the underlying rotation curve, and this contribution may be of the same order of magnitude. However, to a first approximation, the distance uncertainties for different MCs may be considered independent of each other. As can be seen from Table \ref{t:mcs} and Fig.~\ref{f:lands}, the MCs from our sample are distributed so densely along the line of sight within the region potentially illuminated by SS~433 that regardless of their distance uncertainties at least one of these clouds is likely to be X-ray illuminated, unless the SS~433 emission beam is very narrow. 

\begin{figure*}
\includegraphics[bb=30 150 560 700,width=1.0\columnwidth,angle=0]{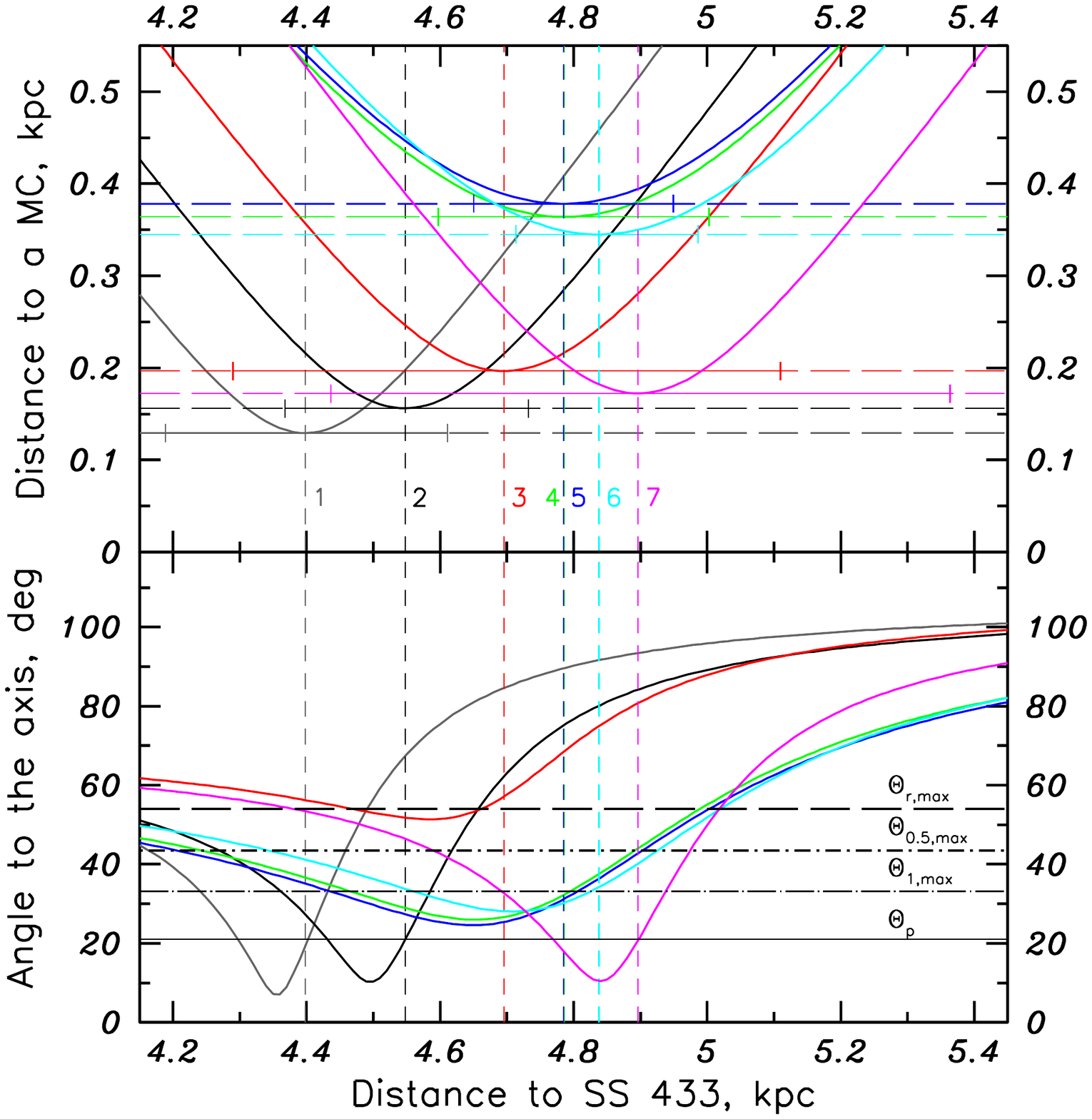}
\includegraphics[bb=30 150 560 700,width=1.0\columnwidth,angle=0]{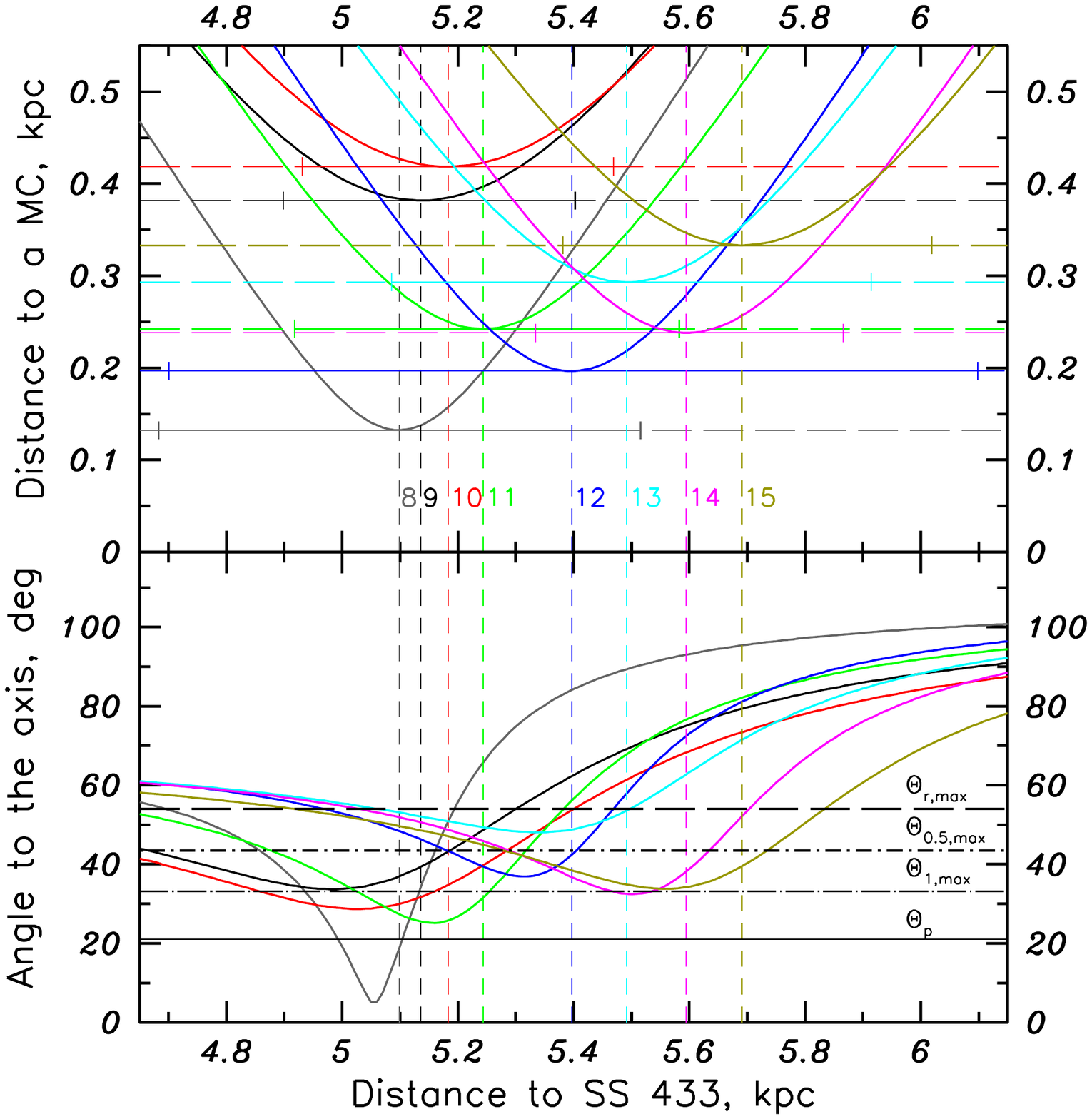}
\caption{Distances from SS 433 (top row) and the corresponding angles
  relative to the precession axis (bottom row) for the 'nearby' (d<5
  kpc, left panels) and 'distant' (d>5 kpc, right panels) molecular
  clouds in Table \ref{t:mcs} as a function of the assumed distance
  from the observer to SS 433. Uncertainties in the distances to the
  MCs (see Table~\ref{t:mcs}) are shown as thin horizontal
  {'errorbars' in the top panels, and they represent the
    uncertainty in the position of the minima of the corresponding
    curves in the top and bottom panels along the x-axis.} Horizontal lines in the bottom panels mark various angles characterising the illuminated region in the case of the widest possible emission beam ($ \Theta_{r}=\Theta_{r,max}=54\deg$). If $ \Theta_r\ll\Theta_p $,  only molecular clouds with $ \theta \approx \Theta_p $ will be illuminated.}
\label{f:mcda}
\end{figure*}

In reality, the situation might be more complicated if some large-scale coherent structures like spiral arms or sites of interaction between supernovae remnants (SNRs) and the molecular gas would cause correlated disturbances in the kinematic distance estimates for the MCs. We have checked Green's catalogue\footnote{https://www.mrao.cam.ac.uk/surveys/snrs/}
\citep{Green2014} and found 6 SNRs whose projected positions lie close to those of MCs in our sample: G36.6-0.7 \citep{Fuerst1987} (close to G036.89-00.41), G038.7-1.3 \citep{Sabin2013} (close to G039.04-00.91), 3C396 \citep{Su2011} (close to G039.34-00.31 and G039.34-00.26), G40.5-0.5\citep{Sun2011} (close to G040.34-00.26), 3C397 \citep{Jiang2010} (close to G041.04-00.26) and G41.5+0.4 \citep{Kaplan2002} (close to G041.24+00.39). All these cases are marked with a '$ {\star} $' symbol in Table~\ref{t:mcs}. Apart from affecting the distance estimates for MCs, the SNRs may also contaminate the X-ray signal from them  (although SNR emission is typically rather soft, with only a small fraction of photons above 4 keV). It is difficult to estimate these potentially important effects without a detailed analysis. 

It should be noted that \cite{Yamamoto2008} previously explored the
molecular gas distribution in the vicinity of the projected position
of SS 433 at somewhat larger scales, within $ |b|<5\deg $ from the
Galactic plane. They found ten {additional} MCs approximately
aligned with the projected direction of the SS 433 jet precession
axis, but with $ V_{LSR} $ ranging from 42.1 to 55.8 km/s, which
corresponds to kinematic distances well outside the range of interest
here {(which is more than 60 km~s$^{-1}$ for all $37\deg\leq l\leq 42\deg $)}. Still, if some clouds in the sample of \cite{Yamamoto2008} are associated with SS 433, the X-ray reflection signal from them can be readily predicted using the same approach as presented in the next section.

\section{Predicted signal}
\label{s:prediction}

Any soft X-ray radiation ($ E\lesssim 1$ keV) emitted by SS~433 and reflected towards us by atomic or molecular gas near the Galactic plane will suffer absorption in the ISM between the source and the scattering site ($ N_{H}\sim hn_{H}\sim 6\times 10^{20}$ cm$^{-2}$), inside the scattering site (if it is a molecular cloud, $ N_{H}\sim r~2n_{H_2}\sim (0.3$--$1.8)\times 10^{22}$ cm$^{-2}$, see Table \ref{t:pred}), and between the scattering site and the observer ($ N_{H}\sim dn_{H}\sim 10^{22}$ cm$ ^{-2} $), the latter value being consistent with the $ N_{H} $ absorption column measured for SS 433 \citep{Medvedev2010,Marshall2013,Khabibullin2016}. To avoid possible uncertainties related to absorption, and taking into account the assumed spectral shape of the primary emission (see below), we will focus only on photons with $E>E_1=4$ keV, for which the effective absorption cross-section $\sigma_{abs}< 10 \sigma_{T} =6.65\times10^{-24}$ cm$ ^2 $ (where $\sigma_T$ is the Thomson cross-section) for gas with solar metallicity \citep{Morrison1983}, so the total optical depth for absorption along the paths of such photons is rather low $\tau_{abs}<0.2 $. 

The luminosity and spectrum of the putative collimated X-ray emission of SS 433 are virtually unknown despite some previous attempts to estimate them indirectly (e.g. \citealt{Panferov1993,Medvedev2010}). We assume that SS 433 'face-on' appearance is similar to ULXs \citep{Fabrika2015}, i.e. the apparent luminosity at energies above 4~keV is $ L_{>4}=L_{39}\times10^{39}$ erg/s with $ L_{39} \sim 1$--10. The X-rays may cause photo-ionization and photo-heating effects in the illuminated gas. However, the relevant ionization parameter {at distance $ R$  from the primary source is given by} (taking into account that most of the soft emission with $E<4$~keV is absorbed between the primary source and the reflector)
\begin{equation}
\xi=\frac{L_{>4}}{nR^2}\sim\frac{1}{400}\frac{L_{39}}{nR_{200}^2},
\label{eq:xi}
\end{equation}
{with $ R_{200}=R/200$ pc $ \sim 1 $, and} it is small even for atomic gas ({with number density} $ n\sim n_{H}\sim 1$ cm$ ^{-3}$), so we may ignore these effects in our treatment below.

In general, the efficiency with which photons from the primary source are reflected to the observer depends on the photon energy and the angle between their initial and final directions. For the geometry considered here (see Section \ref{s:geometry}), the typical scattering angles are close to $ \pi/2 $. In such a case, the cross-section for scattering of X-ray photons by electrons bound in hydrogen atoms or molecules (including the contributions of Rayleigh, Raman and Compton scattering) is approximately equal to the corresponding cross-section for Thomson scattering by free electrons (as far as relativistic corrections can be neglected), i.e.
\begin{equation}
\frac{d\sigma}{d\omega}\approx\left(\frac{d\sigma}{d\omega}\right)_{T,~\pi/2}=\frac{3}{16\pi} \sigma_{T}
\end{equation}
per electron \citep{Sunyaev1996,Sunyaev1999}. This simplifies our further treatment, since the predicted luminosity of the scattered emission (above 4 keV) does not depend on the particular spectral shape of the source. For a viewing direction approximately perpendicular to the initial (before scattering) direction of the emission, the observed reflected flux will be 
\begin{equation}
\label{eq:flucxsc}
f = \frac{3}{16\pi}\frac{\sigma_{T}}{d^2}\frac{L_{>4}}{4\pi R^2}\times N_e ~{\rm erg/s/cm^2},
\end{equation}
where $ R $ is the distance from the primary source to the reflector, $ d $ is the distance from the reflector to the observer and $ N_e $ is the total number of electrons in the reflector. The last quantity can be expressed in terms of the reflector mass $ N_e=M/\mu_e m_p $, where $ \mu_e\approx 1.18 $ is the molecular weight per electron (for solar chemical composition) and $ m_p$ is the proton mass. The \textit{apparent} luminosity of a reflector of mass $ M $ is then
\begin{equation}
L_{sc,>4}=4\pi d^2f=\frac{3}{16\pi}\frac{\sigma_{T}}{R^2}\frac{M}{\mu_e m_p}L_{>4}\approx1.05\times 10^{33} \frac{L_{39}}{R_{200}^2}~{M_4}~{\rm erg/s},
\label{eq:lumsc}
\label{eq:lsc}
\end{equation}
with $ R_{200}=R/200 $ pc and $ M_4=M/10^4M_{\odot}$ (see also \citealt{Cramphorn2002}).

Although the observed spectra of ULXs show some diversity, both from source to source and between spectral states of a given source, their spectral shape above 4 keV can usually be described by a power law with an exponential cutoff at $ E=E_c\sim$few keV (e.g. \citealt{Sazonov2014,Walton2014}). We may thus expect the reflected emission to have a similar spectral shape, except for the presence of an iron fluorescence line at 6.4 keV and absorption edge at $ E_0=7.1 $ keV. The expected equivalent width of the fluorescent line is $\sim 1$ keV with a rather weak dependence on the actual spectral shape (e.g. \citealt{Sunyaev1998}); in the Appendix we make more precise predictions for cutoff-powerlaw models and also calculate flux conversion coefficients from our working energy band (above 4 keV) to more commonly used X-ray energy ranges (e.g. 2--10 keV).

\subsection{Reflection from atomic gas }
\label{ss:atomic}
Let us consider a long bar of atomic gas near the Galactic plane that is parallel to our line of sight and covers a small area $ \Delta S$ on the sky. If this bar contains the point where the SS~433 precession axis crosses the Galactic plane, the amount of the X-ray illuminated gas within the bar  
\begin{equation}
M_{HI}=d_{\ss}^2\Delta S\Delta d~n_{H}m_{p},
\end{equation}
As was shown above, $ \Delta d\approx 2p\approx 170(\tan\Delta\Theta/\tan\Theta_p)d_5$~pc in both weakly collimated and narrow-angle emission regimes.
In what follows, we assume that $ n_{HI}=1 $ cm$ ^{-3} $ in the considered region of the Galactic disc (see Section~\ref{sss:atomic}). 

We may then use equation~(\ref{eq:lumsc}) to determine the X-ray luminosity per square degree arising due to reflection by the atomic ISM in the region of the Galactic plane ($b=0\deg$) illuminated by SS~433:
\begin{equation}
L_{HI,>4}\approx 2.9\times10^{33}~k {L_{39}}~d_5 ~{\rm erg/s/deg^2},
\end{equation}
where we have replaced $ R_{200} $ with $ R_0/200 $pc$=1.085~d_5 $ and introduced
\begin{equation}
k= \frac{\tan\Delta\Theta}{\tan\Theta_p}.
\label{eq:k}
\end{equation}

The corresponding X-ray surface brightness
 \begin{equation}
 \label{eq:shi}
  S_{HI,>4} (0) \approx 9.6 \times 10^{-13} k\frac{L_{39}}{d_5}~{\rm erg/s/cm^2 /deg^2 }.
 \end{equation}

It is easy to show that the brightness of the reflected signal decreases with Galactic latitude as
\begin{equation}
S_{HI,>4}(b)=\frac{\exp(-|b|/b_d)}{|b/b_{\rm \ss}-1|}~S_{HI,>4}(0),
\end{equation} 
where $ b_d\approx z_d/d_{\ss}\approx 0.46\times d_{5}^{-1} $ deg and we recall that $z_d=40$~pc is the scale height of the Galactic atomic gas. The average surface brightness within $ \Delta b \ll b_{\rm \ss}$ of the Galactic plane is thus
\begin{equation}
<S_{HI,>4}(\Delta b)>=S_{HI,>4} (0)(1-\exp^{-\Delta b/b_d})\frac{b_d}{\Delta b}.
\end{equation}
For $ \Delta b=0.5 $ deg, this yields  
\begin{equation}
<S_{HI,>4}>_{0.5}\approx 0.6~S_{HI,>4} (0)\approx 5.8 \times 10^{-13}~{k}\frac{L_{39}}{d_5}~{\rm erg/s/cm^2 /deg^2 }.
\label{eq:shiaverage}
\end{equation}

\subsection{Reflection from molecular clouds}
\label{ss:molecular}
\begin{table*}
\caption{Physical parameters of the relevant molecular clouds and the predicted reflection signal from them. The first two columns are identical to those in Table \ref{t:mcs}. The third column {(R$_{min}$)} gives the minimal distance from SS 433 to a given cloud divided by the characteristic scale 200 pc (see Fig. \ref{f:mcda}), followed by columns with estimates for the cloud's size {(r)}, density {(n$_{\rm H_2} $)}, area {(Area)}, mass {(M)} and its uncertainty {($ \delta$M)} as provided by the corresponding catalogue entries in \citep{RD2010}. The next two columns {(S$ _{sc,200} $ and f$ _{sc,200} $)} give the expected surface brightness and total flux of reflected X-ray emission from the cloud for assumed apparent luminosity of SS~433 $ L_{4-10}=10^{39} $ erg/s, {distance to it $ d_{\ss}=5 $ kpc and distance from SS 433 to the cloud} $ R=200 $ pc. The last column {(f$_{rxte} $)} presents upper limits on the reflected flux obtained from the $ RXTE $ map (see text), where values marked by $ ^\dagger $ are likely affected by a nearby bright source. All fluxes are in the 4--10~keV energy range.}
\begin{tabular}{c|c|c|c|c|c|c|c|c|c|c|c}
\hline
  \multicolumn{1}{|c|}{} &
  \multicolumn{1}{|c|}{GRS } &
  \multicolumn{1}{c|}{{R$_{\rm min}$}} &
    \multicolumn{1}{c|}{r} &
      \multicolumn{1}{c|}{n$ _{\rm H_2} $  } &
  \multicolumn{1}{c|}{Area} &
  \multicolumn{1}{c|}{M} &
  \multicolumn{1}{c|}{$ \delta$M} &
    \multicolumn{1}{c|}{S$ _{sc,200} $/$10^{-11}$ } &
  \multicolumn{1}{c|}{f$ _{sc,200} $/$10^{-11}$ } &
  \multicolumn{1}{c|}{f$_{rxte} $/$10^{-11}$ } \\
    \multicolumn{1}{|c|}{} &
  \multicolumn{1}{|c|}{name } &
  \multicolumn{1}{c|}{200 pc} &
    \multicolumn{1}{c|}{pc} &
  \multicolumn{1}{c|}{cm$^{-3}$} &
  \multicolumn{1}{c|}{deg$ ^2 $} &
  \multicolumn{1}{c|}{$ 10^{4}$M$_ \odot $} &
  \multicolumn{1}{c|}{$ 10^{4}$M$_ \odot $ } &
    \multicolumn{1}{c|}{erg/s/cm$ ^2$/deg$ ^2 $}&
  \multicolumn{1}{c|}{erg/s/cm$ ^2$}&
  \multicolumn{1}{c|}{erg/s/cm$ ^2$} \\  
\hline
1 & G039.29-00.61 & 0.65 & 6.7 & 194.2 & 0.076 & 1.6 & 0.6 &{0.7}&
 0.05 & 0.8 \\
2 & $^\bigstar $G039.34-00.31 & 0.78 & 7.7 & 200.9 & 0.14 & 2.5 & 1.0 & {0.6}&
0.08 &0.8 \\
3 & $^\bigstar $G041.04-00.26 & 0.98 & 25.6 & 94.6 & 0.12 & 44.0 & 7.0 & {12.0}&
1.5 & 3.0 $^\dagger$\\
4 & G036.44+00.64 & 1.82 & 6.7 & 436.2 & 0.026 & 3.6 & 1.0 & {4.6}&
0.1 & 10$^\dagger$\\
5 & G036.39+00.84 & 1.89 & 5.0 & 351.0 & 0.06 & 1.2 & 0.4 & {0.67}&
0.04 &1.3\\
6 & G036.54+00.34 & 1.73 & 1.6 & 338.6 & 0.074 & 0.041 & 0.01&{0.02}&
  0.001 & 10$^\dagger$\\
   \smallskip
7 & $^\bigstar $G039.34-00.26 & 0.86 & 9.0 & 173.0 & 0.45 & 3.4 & 1.0 & {0.24}&
0.1 & 0.8
\\
8 & $^\bigstar $G039.04-00.91 & 0.66 & 8.2 & 194.8 & 0.26 & 2.9 & 1.0 & {0.37}&
0.1 & 1.0\\
9 & G036.14+00.09 & 1.91 & 2.8 & 288.8 & 0.025 & 0.17 & 0.07 &{0.23}&
 0.006 &12.2$^\dagger$ \\
10 & G036.09+00.64 & 2.09 & 9.8 & 214.1 & 0.073 & 5.4 & 2.0 &{2.5}&
 0.2 &7.1$^\dagger$
 \\
11 & G037.74-00.46 & 1.21 & 8.8 & 95.3 & 0.3 & 1.8 & 0.7 & {0.2}&
0.06 & 3.4$^\dagger$
\\
12 & $^\bigstar $G040.34-00.26 & 0.98 & 10.5 & 160.6 & 0.14 & 5.1 & 1.0 & {1.2}&
0.2 & 1.1\\
13 & $^\bigstar $G041.24+00.39 & 1.47 & 3.6 & 221.3 & 0.3 & 0.28 & 0.1 & {0.03}&
0.009 &1.1\\
14 & G037.69-00.86 & 1.19 & 5.6 & 195.5 & 0.19 & 0.95 & 0.4 & {0.17}&
0.03 &1.1 \\
15 & $^\bigstar $G036.89-00.41 & 1.67 & 16.0 & 110.2 & 0.12 & 12.0 & 4.0 & {3.3}& 
0.4 &20$^\dagger$
 \\
\hline\end{tabular}

\label{t:pred}
\end{table*}

Using equation~(\ref{eq:flucxsc}), we can evaluate the expected X-ray reflected flux from a molecular cloud:
\begin{equation}
\label{eq:fh2sc}
f_{H_2,>4}=3.5\times 10^{-13} \frac{L_{39}}{d_5^2 R_{200}^2}{M_4}~{\rm erg/s/cm^2}.
\end{equation} 
The corresponding X-ray surface brightness is 
\begin{equation}
S_{H_2,>4}= \frac{f_{H_2,>4}}{\pi\delta {\rm l} \delta {\rm b}}\approx 1.1\times 10^{-13}\frac{L_{39}}{d_5^2 R_{200}^2} \frac{M_4}{\delta {\rm l} \delta {\rm b}}~{\rm erg/s/cm^2/deg^2},
\end{equation}
assuming that the projection of the MC on the sky has an elliptical shape with major axes $ 2\delta {\rm l} $ and $ 2\delta {\rm b} $ (see Section \ref{sss:molecular}).

Table \ref{t:pred} gives the predicted fluxes for the MCs in our sample for some characteristic values of $ R=200 $ pc and $ d_{\ss}=5 $ kpc, and assuming that the whole cloud is located inside the X-ray beam, which is a reasonable assumption since the typical size of our MCs is $ \sim 10 $ pc. The actual distance $ R $ from SS 433 to the reflector is determined by the true distances of SS 433 and a given MC from the observer (see the top panels in Fig.~\ref{f:mcda}). In addition, one should apply a correction for the fraction of time a given direction spends inside the illumination cone, which is given by the function $ F(\theta)$ of the angle between the precession axis and the line connecting the source and a given MC (see Section \ref{sss:precession}). This angle is also determined by the actual distances of SS 433 and the MCs (see the bottom panels in Fig.~\ref{f:mcda}). For the limiting case $ \Theta_r\ll\Theta_p$, only those clouds for which the angle curve in the bottom panels of Fig.~\ref{f:mcda} approaches $ \Theta_p $ close enough (i.e. $ |\theta-\Theta_p|<\eta\Theta_{p} $) can be illuminated. As is obvious from Fig.~\ref{f:mcda}, only four clouds (G039.29-00.61, G039.34-00.31, G039.34-00.26 and G039.04-00.91, {i.e. clouds 1, 2, 7 and 8 in Tables \ref{t:mcs} and \ref{t:pred}} ) have a chance to be illuminated in this narrow-beam regime.

\section{Constraints from \textit{RXTE} and \textit{ASCA} data}
\label{s:constraints}
\begin{figure*}
\centering{
\includegraphics[bb=30 0 700 579,width=1.45\columnwidth,angle=0]{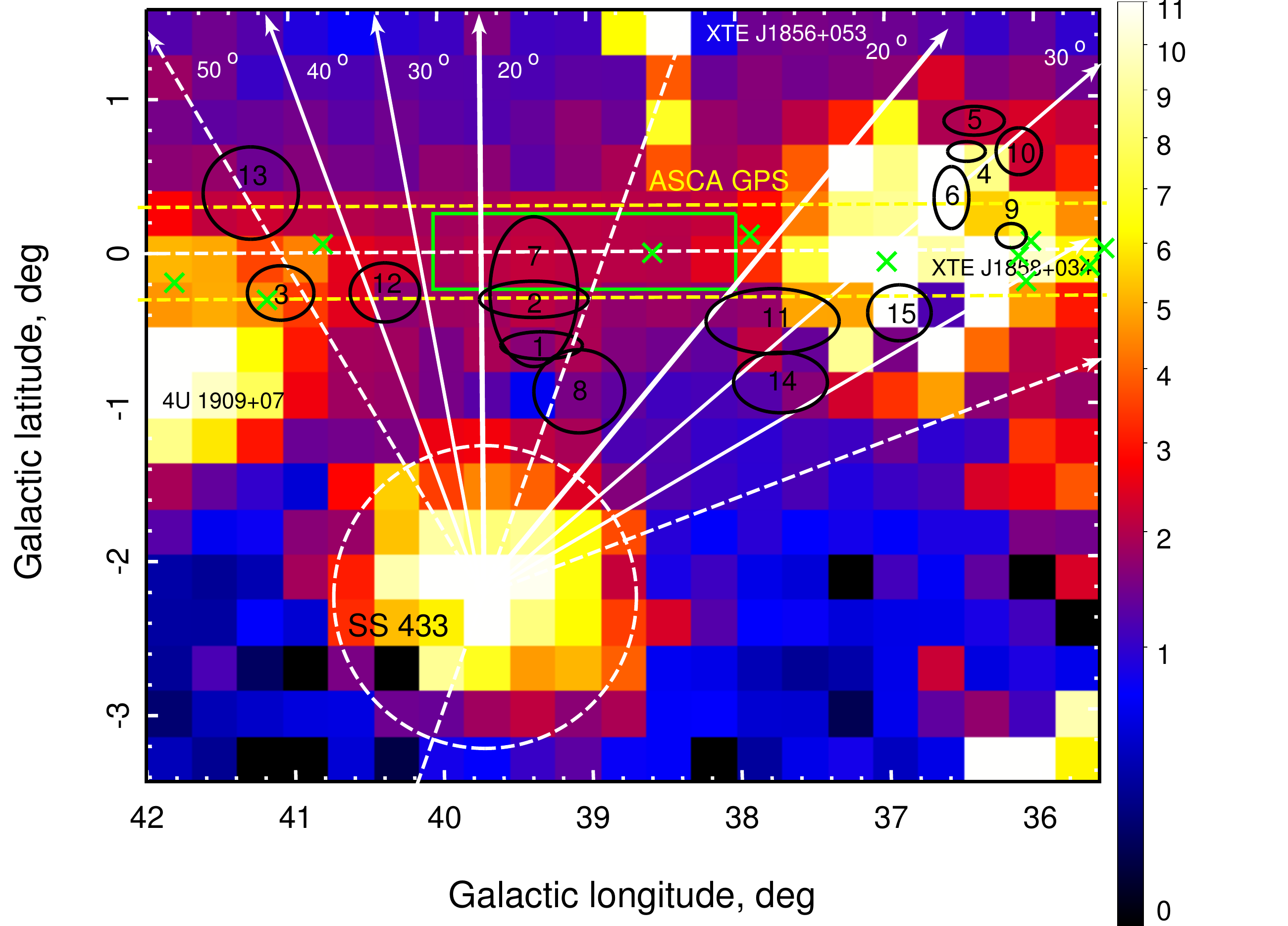}
\caption{The region of the sky where 'X-ray echos' of SS 433 may be found. The background image is based on the \textit{RXTE} count rate map (cnts/s per pixel) in the 3--20 keV energy range (with pixel size 0.3 deg) by \citealt{Revnivtsev2006}. Yellow lines outline the $|b|\leq 0.3\deg $ region of the \textit{ASCA} Galactic plane survey in the 0.7--10 keV energy range, while green crosses mark the positions of sources detected in this survey \citep{Sugizaki2001}. Black ellipses show the positions and approximate shapes of the molecular clouds in the vicinity of SS 433 ($d\sim 5\pm 1 $ kpc), as provided by the BU-FCRAO Galactic Ring Survey in $^{13}$CO  $J = 1-0$ line \citep{RD2009}. The numbers above the ellipses show the estimated distances to them. The white circle around SS 433 is 2 degrees in diameter, which approximately corresponds to the extent of the W50 nebula in the direction of the jet  precession axis, shown with the central dashed line. The other white lines depict projections of cones with various opening angles (as labeled) and the axis coinciding with the precession axis. The green rectangular 2$\deg \times $ 0.5$ \deg $ region with the center at b=$0\deg $ is used for extraction of the average \textit{RXTE} count rate for setting an upper limit on the reflected signal (see text).}
\label{f:xtemap}
}
\end{figure*}
Below we obtain some constraints on the collimated X-ray emission of SS 433 by comparing the predictions of the previous section with existing X-ray data for the sky region under consideration. Since the expected equivalent width of the Fe I K$ \alpha $ fluorescent line arising due to reflection is comparable to the equivalent width of the Fe-lines complex in the spectrum of the Galactic X-ray ridge emission ( $ \sim 1 $ keV) \citep{Revnivtsev2006line}, searching for a reflected signal in a narrow band around the 6.4~keV line would not provide any significant advantage (with the current generation of X-ray telescopes and instruments), except for avoiding contamination from the brightest nearby sources \citep{Revnivtsev2006line}. Hence, we restrict our consideration to data from the \textit{ASCA} and \textit{RXTE} observatories obtained in the broad energy ranges of 4--10 keV and 3--20 keV, respectively.     
\subsection{Atomic gas}
\label{ss:atomicconstr}
The \textit{ASCA} Galactic Plane Survey covered the stripe $ |b|<0.3\deg $ along the Galactic plane (see Fig.~\ref{f:xtemap}) with the mean exposure time of $ \sim $ 20 ks \citep{Sugizaki2001} and is well suited for measuring the X-ray surface brightness along the Galactic plane. 

After subtraction of the cosmic X-ray background (CXB) contribution, $S_{CXB,4-10}\approx 1\times 10^{-11}$ erg/s/cm$^2$/deg$^2$ {\citep{Sugizaki2001}}, the measured surface brightness in the 4--10 keV energy band averaged over the $ 38\deg<{\rm l}<40\deg $ region is $S_{4-10}\approx 1.2 \times 10^{-11}$ erg/s/cm$^2$/deg$^2$  \citep{Sugizaki2001}. \textit{RXTE} data in the 3--20 keV range yield an estimate $S_{3-20}\approx 2.6\times 10^{-11}$ erg/s/cm$^2$/deg$^2$ (also CXB subtracted) for the same region (depicted as a green box in Fig.~\ref{f:xtemap}), which is consistent with the expected ratio $S_{3-20}/S_{4-10}\approx 2$ for the $ \Gamma\approx 2 $ spectrum of the GRXE.

{ However, the well-established correlation} between the infrared (cumulative starlight) and X-ray surface brightness (cumulative emission of unresolved point X-ray sources) predicts a somewhat lower brightness of the GRXE  emission in this region: $2.2\times 10^{-11}$ erg/s/cm$^2$/deg$^2$ for 3--20 keV \citep{Revnivtsev2006}, and hence $ 1.1\times 10^{-11} $ erg/s/cm$^2$/deg$^2$ for 4--10 keV. Taking into account that the IR-X-ray correlation allows a scatter of $\sim 20\% $ \citep{Revnivtsev2006}, we conclude that the surface brightness of the non-GRXE emission in this region is between zero and $ \sim 8\times 10^{-12} $ erg/s/cm$^2$/deg$^2$ in the 3--20 keV energy range, or $ \sim 4\times 10^{-12} $ erg/s/cm$^2$/deg$^2$ in the 4--10 keV range. 

Assuming that all of this residual signal (if any) is due to scattered collimated X-ray emission of SS 433, we may set a constraint on its apparent  luminosity in the 4--10 keV energy range, using equation~(\ref{eq:shiaverage}):
\begin{equation}
  L_{4-10}< 7 \times 10^{39} ~ \frac{d_5}{k}~{\rm erg/s}.
\label{eq:aconstrain}
\end{equation}
This result can be converted to a different energy band using the conversion factors for a given spectral shape given in the Appendix. For the standard X-ray band of 2--10 keV, the conversion factor is $ \approx 2$ for a power-law spectrum with $ \Gamma=2$ and a cutoff above 3 keV, hence the upper limit is
\begin{equation}
  L_{2-10}<  1.4\times 10^{40}~ \frac{d_5}{k}~{\rm erg/s}.
 \label{eq:aconstrain2}
\end{equation}


\subsection{Molecular clouds}
\label{ss:molecularconstr}

For the MCs in our sample, the coverage of \textit{ASCA} GPS is not sufficient, since most of the clouds do not fall in the $ |b|<0.3\deg $ stripe (see Fig.~\ref{f:xtemap}). The RXTE flux map in the 3--20 keV range with its 0.3 deg {pixel size} (see Fig.~\ref{f:xtemap}, \citealt{Revnivtsev2006}), is not very  suitable in this case either, since, although the predicted X-ray surface brightness for MCs may be rather high, their projected area $ S_{MC} $ is typically small (see Table \ref{t:pred}) compared to the field of view of the \textit{RXTE}/PCA (non-imaging) instrument, $S_{FOV}\sim 1$ deg$^2 $. As a result, the signal from the MCs is expected to be smeared out by a factor of $ S_{FOV}/S_{MC} $, making them hardly detectable against the background of the extended ridge emission (having a characteristic scale-height of $ \sim 1 \deg$) and the contaminating signal from bright nearby sources (see Fig.~\ref{f:xtemap}). 

Adopting a conversion factor from count rates to 3--20 keV flux for point sources of 1 cts/s $\approx 2 \times 10^{-11} $ erg/s/cm$ ^2 $ (this corresponds to a power-law spectrum with $ \Gamma=2 $ and the slew observational mode, in which the majority of observations contributing to the \textit{RXTE} map were performed, \citealt{Revnivtsev2004,Sazonov2008}), we estimated flux limits for non-GRXE emission from the MCs in our sample regarding them as point sources. Specifically, we used the count rate in the brightest pixel within the MC ellipse and subtracted from it the expected GRXE contribution at this position, taking into account that the GRXE brightness declines exponentially with latitude with a scale-height of $\approx 1.8 \deg$ at $ l\sim40\deg $ and offset from the Galactic plane of $ \Delta b=-0.15\deg $ (as follows from the model presented in \citealt{Revnivtsev2006}). In order to get an upper limit on the residual flux, we actually reduced the expected contribution of the GRXE by 20\% to take into account the uncertainty in the IR-X-ray correlation (see above).

Taking into account the systematic uncertainties involved in the procedure of GRXE subtraction from the \textit{RXTE} map, we see no significant excesses from any MC in our sample, except for some clouds in the projected vicinity (namely within $\sim 1\deg$, corresponding to the radius of the \textit{RXTE} FOV) of bright X-ray sources (see Fig.~\ref{f:xtemap}). Table \ref{t:pred} lists the resulting 4--10 keV flux limits on the reflected emission from the MCs, derived from the corresponding 3--20 keV values using the constant conversion factor $\approx 2 $ (see Appendix). Depending on the actual distances between SS~433 and the MCs (and hence the angle at which a given MC is 'viewed' from SS 433 relative to the precession axis) and the half-opening angle of the emission cone, these limits may or may not actually provide constraints on the luminosity of SS~433.

If the SS~433 emission cone is wide ($\Theta_r>\Theta_p$), the probability that none of the MCs fall inside the illuminated region is rather low. Furthermore, the four clouds (G039.04-00.91, G039.29-00.61, G039.34-00.31 and G039.34-00.26) {that (based on their projection on the sky close to the projection of the precession axis, see Fig. \ref{f:xtemap}) can be located within the precession cone,} are fairly likely to fall inside the region of permanent illumination ($ \theta<\Theta_1 $), and the ratio $ f_{xte}/f_{sc} $ varies from 8 to 16 for them (see Table \ref{t:pred}). However, even for these clouds, the lowest possible upper limit on the luminosity is $ L_{4-10}\sim 8\times 10^{39}$ erg/s, or $ L_{2-10}\sim 1.6\times 10^{40}$ erg/s. Therefore, with the available \textit{RXTE} data, MCs do not provide better constraints on the SS~433 luminosity compared to the atomic gas near the Galactic plane.

In the regime of strong collimation ($\Theta_r<\Theta_p$), the situation becomes even more strongly contingent on the relative positions of the MCs and SS 433, since only those clouds 'viewed' from SS 433 at an angle close to $ \Theta_p$ relative to the precession axis (see Section \ref{ss:molecular} and Fig.~\ref{f:mcda}) can reflect its emission. Here again, the four clouds G039.04-00.91, G039.29-00.61, G039.34-00.31 and G039.34-00.26 have the highest chance to be illuminated, but we should recall that in the narrow-beam regime any given direction is illuminated only a small fraction of time $ \approx \eta/\pi $, which diminishes the reflected signal by the same factor.

The constraints obtained from reflection by MCs can be improved using X-ray observations with better angular resolution than \textit{RXTE}. In fact, some of the relevant clouds have been serendipitously observed by \textit{Chandra}, \textit{XMM-Newton}, \textit{Suzaku} or \textit{Swift}/XRT. Although the small FOV of these instruments is usually not sufficient for covering the whole MC, their high angular resolution makes it possible to resolve the small-scale structure of the cloud, which might provide some advantage for detection of X-ray reflected emission. We plan to undertake such an analysis in future work. 

\section{Summary and discussion}
\label{s:discussion}


We can summarise the luminosity constraints imposed by the absence of significant reflected emission from the atomic gas in the Galactic plane (eq.~\ref{eq:aconstrain2} in the previous session) as follows. If the X-ray emission cones in SS~433 are not very wide, namely $\Theta_r<38\deg$, then the geometrical coefficient $ k\gtrsim 0.5(\Theta_r/\Theta_p)^2$ (see Fig.~\ref{f:frac} and eq.~\ref{eq:k}) and the upper limit on the SS~433 apparent (i.e. isotropic-equivalent) luminosity is
\begin{equation}
  L_{2-10}< 2.8\times 10^{40} d_5\left(\frac{\Theta_p}{\Theta_r}\right)^2~{\rm erg/s},
\end{equation}
while the corresponding limit on the angular-integrated luminosity
\begin{equation}
 L_{c,2-10}\approx 0.5\Theta_r^2 L_{2-10}< 1.8\times 10^{39}~{d_5}~{\rm erg/s}.
\end{equation}

For large angles of collimation, $38\deg<\Theta_r<54\deg$ (where $54\deg$ is the largest possible angle consistent with the fact that SS~433 collimated X-ray emission is not observed directly at any precession or nutation phase), $k\approx \tan\Theta_r/\tan\Theta_p$ and the corresponding constraints are
\begin{equation}
  L_{2-10}< 5.5\times 10^{39}~{d_5}(\tan\Theta_r)^{-1} ~{\rm erg/s}
\end{equation}
and
\begin{equation}
L_{c,2-10}< 1.65 \times 10^{39}~{d_5}~[1+0.4(\Theta_r-\pi/4)]~{\rm erg/s}.
\end{equation}

Therefore, we have obtained a robust upper limit of $\sim 2\times 10^{39}$~erg~s$^{-1}$ on the total luminosity of the putative collimated X-ray emission of SS~433, which is only weakly dependent on the opening angle of the radiation cone. In contrast, the upper limit on the SS~433 apparent luminosity monotonically decreases with increasing $\Theta_r$ and is $L_{2-10}\lesssim 3\times 10^{40}$~erg~s$^{-1}$ for $\Theta_r\gtrsim\Theta_p=21\deg$. It is the apparent luminosity that would be perceived by an observer looking at the SS~433 super-critical accretion disk face-on. 

The limits imposed by the absence of a significant X-ray reflection signal from the molecular clouds near SS~433 are currently somewhat weaker than those provided by the atomic gas, but could be improved with higher angular resolution X-ray observations. Improvements in the search for scattered X-ray emission from the atomic gas are also possible in the future. In particular, \textit{ASTRO-H}, with its excellent energy resolution, could provide maps of the Galactic plane region under consideration { ($ 38\deg\leq ~l~\leq 40\deg $)} in the expected iron fluorescent line at 6.4~keV distinguishing it from the dominant 6.7~keV line in the GRXE. We also note that the scattered emission discussed here should be nearly 100\% polarised since we are dealing with scattering at almost right angles (e.g. \citealt{Churazov2002}), while the competing sources of extended emission such as the GRXE (consisting of numerous point sources) are unlikely to produce any noticeable polarisation. Therefore, the reflected signal from SS~433 might be an interesting target for future X-ray polarisation missions. 

The limit on the apparent luminosity derived here is still consistent with SS~433 being a misaligned ULX \citep{Fabrika2015}, since typical observed (2--10~keV) luminosities of such objects are a~few~$10^{39}$ to $\sim 10^{40}$~erg~s$^{-1}$. However, this limit is stringent enough to rule out that SS~433 is as X-ray luminous as the brightest known ULXs (with luminosities higher than $\sim 10^{41}$~erg~s$^{-1}$) \citep{Feng2011}, unless its emission beam is very narrow, $\Theta_r\ll\Theta_p=21\deg$. 

{
It should be noted that the limits obtained here actually pertain to the luminosity of SS~433 in a relatively recent past ($\sim 500$--1000~years ago), since the reflected emission reaches the observer with a delay of $\sim 200~{\rm pc}/c$ relative to the direct signal from the central regions of the SS~433 jets. However, as was discussed in Section~\ref{ss:source}, the stable behaviour of the jets on $\sim 100$~pc scales implies that the system has been in the same state of activity at least over the past $\sim 1000$~years. Therefore, the luminosity limits obtained here are likely relevant for the current luminosity of SS~433 as well.
}

We have actually constrained only the luminosity above 4 keV. Unfortunately, softer emission from SS~433, especially at energies below 2 keV, cannot be constrained by the same method, because of significant interstellar absorption both between SS~433 and the potential scatterers (atomic gas and molecular clouds) and between the scatterers and us. In fact, there is a class of sources, called ultraluminous supersoft sources (ULSs), which have supersoft spectra (with almost no photons above 2 keV) and apparent bolometric luminosities of a few  $ 10^{39} $ erg~s$^{-1}$. It has been suggested (see \citealt{Urquhart2015} and references therein) that such objects might be another variety of supercritically accreting stellar mass black holes or neutron stars, with the distinction between ULXs and ULSs arising due do differences in the specific mass accretion rate $\dot{m} $ and/or viewing angle of the observer,  with ULSs corresponding to higher $\dot{m} $ for a fixed viewing angle and larger viewing angles for a fixed $\dot{m}  $ \citep{Urquhart2015}. It is possible, especially given its very high accretion rate $ \dot{m}\sim 400$ \citep{Fabrika2004}, that SS~433 is an ULS. Moreover, very recently \cite{Liu2015} discovered precessing baryonic jets in one of the ULSs, which makes the possible association of such sources with SS~433 even more intriguing. If there is indeed luminous and collimated supersoft emission from SS~433, it might be revealed through its influence on the thermal and ionization stage of the plasma in the closer vicinity of the source (i.e. inside the W50 nebula) and also, perhaps, on the plasma in some parts of the jets. Such future studies could help constrain the SS 433 luminosity across a broader X-ray energy range.


\section*{Acknowledgements} 
{ We are grateful to Mikhail Revnivtsev for providing the
  \textit{RXTE} map and assisting with its analysis.} The research was
supported by the Russian Science Foundation (grant 14-12-01315). IK is
grateful to the Max-Planck-Institut f\"ur Astrophysik for hospitality
and the Dynasty Foundation for the fellowship support. {We are
  grateful to the referee for useful comments and suggestions.}

\section*{Appendix}
\label{s:app}
Suppose that the photon luminosity of the illuminating source is 
\begin{equation}
I(E)=A~\left(\frac{E}{E_c}\right)^{-\Gamma}~e^{-E/E_c}~{\rm photons~s^{-1}~keV^{-1} },
\end{equation}
where $ A $ is the normalisation constant, which can be determined from the total luminosity above $ E_1=4 $ keV
\begin{equation}
L_{>4}=\int_{4}^{\infty}E I(E) dE ~ {\rm erg~s^{-1}}.
\end{equation}

The expected photon luminosity in the 6.4~keV fluorescent line is
\begin{equation}
L_{6.4}=\frac{\rm Y}{4\pi~R^2}{ N_{Fe}}
\int_{E_0}^{\infty} I(E)\sigma_{Fe}(E) {\rm d}E ~{\rm photons}~ {\rm s}^{-1}~{\rm cm}^{-2}
\end{equation}
for reflection from $ N_{Fe}=\delta_{Fe}\simeq 3\times 10^{-5} N_e$ iron atoms {(assuming the solar metallicity)} located at distance $ R $ from the primary source.

The photoabsorption cross-section entering the above equation is given be  
\begin{equation}
\sigma_{Fe}=\sigma_0 \left(\frac{E}{E_0}\right)^{-3} ~ {\rm cm^2},
\end{equation}
with $ E_0=7.1 {\rm keV} $ and $ \sigma_0=3.53\times 10^{-20}~{\rm cm^2}$, while ${\rm Y}\approx 0.3$ is the Fe K$ \alpha $ fluorescent yield \citep{Sunyaev1998}.

It is convenient to rewrite $ L_4$ and $ L_{6.4} $ as 
\begin{equation}
L_{>4}=A E_0^2\varepsilon^\Gamma \int_{E_1/E_0}^{\infty}{\rm d}\xi\frac{e^{-\xi/\varepsilon}}{\xi^{\Gamma-1}}~{\rm erg~s^{-1}},
\end{equation}
and
\begin{equation}
L_{sc,6.4} ={\rm Y}\frac{4\delta_{Fe}\sigma_0}{3\sigma_T}\frac{{3}\sigma_T N_{e}}{{16\pi} R^2}~A~E_0\varepsilon^\Gamma \int_{E_1/E_0}^{\infty}{\rm d}\xi\frac{e^{-\xi/\varepsilon}}{ \xi^{\Gamma+3}}  ~{\rm erg~s^{-1} keV^{-1}},
\end{equation}
with  $ \varepsilon=E_c/E_0 $.

Then,
\begin{equation}
\frac{E_{0}^2~L_{sc,6.4}}{L_{sc,>4}}={\rm Y}\frac{4\delta_{Fe}\sigma_0}{3\sigma_T}~\varphi_{\Gamma}(E_c)~ {\rm keV},
\label{eq:ew}
\end{equation}
where
\begin{equation}
\varphi_{\Gamma}(E_c)={E_0}\int_{1}^{\infty}{\rm d}\xi\frac{e^{-\xi/\varepsilon}}{ \xi^{\Gamma+3}}/\int_{E_1/E_0}^{\infty}{\rm d}\xi\frac{e^{-\xi/\varepsilon}}{\xi^{\Gamma-1}}~{\rm keV}.
\label{eq:phi}
\end{equation}

\begin{figure}
\includegraphics[bb=60 190 600 670, width=1.1\columnwidth,angle=0]{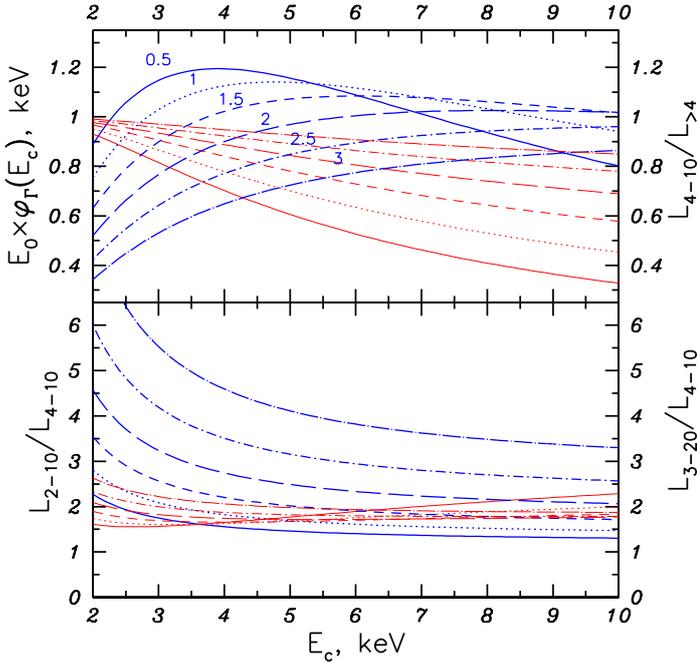}
\caption{\textit{Top panel.} Function $ \varphi_{\Gamma}(E_c)$ (blue curves), which determines the equivalent width of the Fe I K$ \alpha $ line in the scattered emission for spectra of incident radiation described by a power law with an exponential cutoff $E_c  $ and the slope indicated next to each curve. The red curves show the ratio of the luminosities in the hard \textit{ASCA} range (4--10 keV) to that above 4 keV for the same spectra and with the same line-type coding of the curves as for $ \varphi_{\Gamma}(E_c)$. \textit{Bottom panel.} Ratios of the luminosities in the standard X-ray range (2--10 keV, blue curves) and the \textit{RXTE} range (3--20 keV, red curves) to the luminosity from 4 to 10 keV for the same spectra and with the same line-type coding as above.}
\label{f:fluor}
\end{figure}

The quantity on the right-hand side of equation~(\ref{eq:ew}) is approximately equal to the equivalent width of the 6.4 keV line, and since $ {\rm Y} \frac{4\delta_{Fe}\sigma_0}{3\sigma_T}\approx 0.7$, and $\varphi_{\Gamma}(E_c)\approx 1.$ (see the top panel of Fig.~\ref{f:fluor}), one may expect $ EW_{6.4}\sim 0.7 $ keV for the spectra considered here. This formula allows one to predict the flux in the Fe I K$ \alpha $ line for a given scattered flux above 4 keV (i.e. $ L_{sc,>4} $) and assuming some spectral shape of the incident emission. The bottom panel of Fig.~\ref{f:fluor} shows the conversion coefficients between the luminosity above 4~keV and the luminosity in a number of other ranges (namely, 2--10, 4--10 and 3--20 keV).

\begin{thebibliography}{99}

\bibitem[\protect\citeauthoryear{Begelman, King, 
\& Pringle}{2006}]{Begelman2006} Begelman M.~C., King A.~R., Pringle J.~E., 2006, MNRAS, 370, 399 

\bibitem[\protect\citeauthoryear{Benjamin et 
al.}{2005}]{Benjamin2005} Benjamin R.~A., et al., 2005, ApJ, 630, 
L149 

\bibitem[\protect\citeauthoryear{Blundell
\& Bowler}{2004}]{Blundell2004} Blundell K.~M., Bowler M.~G., 2004, ApJ, 616, L159 

\bibitem[\protect\citeauthoryear{Brinkmann, Aschenbach, 
\& Kawai}{1996}]{Brinkmann1996} Brinkmann W., Aschenbach B., Kawai N., 1996, A\&A, 312, 306 

\bibitem[\protect\citeauthoryear{Churazov, Sunyaev, 
\& Sazonov}{2002}]{Churazov2002} Churazov E., Sunyaev R., Sazonov S., 2002, MNRAS, 330, 817 

\bibitem[\protect\citeauthoryear{Cramphorn 
\& Sunyaev}{2002}]{Cramphorn2002} Cramphorn C.~K., Sunyaev R.~A., 2002, A\&A, 389, 252 

\bibitem[\protect\citeauthoryear{Dehnen 
\& Binney}{1998}]{Dehnen1998} Dehnen W., Binney J., 1998, MNRAS, 294, 429 


\bibitem[\protect\citeauthoryear{Dubner et al.}{1998}]{Dubner1998} 
Dubner G.~M., Holdaway M., Goss W.~M., Mirabel I.~F., 1998, AJ, 116, 1842 


\bibitem[\protect\citeauthoryear{Eikenberry et 
al.}{2001}]{Eikenberry2001} Eikenberry S.~S., Cameron P.~B., Fierce B.~W., Kull D.~M., Dror D.~H., Houck J.~R., Margon B., 2001, ApJ, 561, 1027 

\bibitem[\protect\citeauthoryear{Fabbiano}{2006}]{Fabbiano2006} Fabbiano G., 2006, ARA\&A, 44, 323 

\bibitem[\protect\citeauthoryear{Fabrika 
\& Mescheryakov}{2001}]{Fabrika2001} Fabrika S., Mescheryakov A., 2001, IAUS, 205, 268 

\bibitem[\protect\citeauthoryear{Fabrika}{2004}]{Fabrika2004} 
Fabrika S., 2004, ASPRv, 12, 1 
%
\bibitem[\protect\citeauthoryear{Fabrika et 
al.}{2015}]{Fabrika2015} Fabrika S., Ueda Y., Vinokurov A., 
Sholukhova O., Shidatsu M., 2015, Nature Physics, 11, 551 


\bibitem[\protect\citeauthoryear{Feng 
\& Soria}{2011}]{Feng2011} Feng H., Soria R., 2011, NewAR, 55, 166 


\bibitem[\protect\citeauthoryear{Fuerst et 
al.}{1987}]{Fuerst1987} Fuerst E., Reich W., Reich P., Handa T., Sofue Y., 1987, A\&AS, 69, 403 


\bibitem[\protect\citeauthoryear{Gilfanov}{2004}]{Gilfanov2004} 
Gilfanov M., 2004, MNRAS, 349, 146 

\bibitem[\protect\citeauthoryear{Gilfanov, Grimm, 
\& Sunyaev}{2004}]{GilfanovGS2004} Gilfanov M., Grimm H.-J., Sunyaev R., 2004, MNRAS, 351, 1365 


\bibitem[\protect\citeauthoryear{Green}{2014}]{Green2014} Green D.~A., 2014, BASI, 42, 47 

\bibitem[\protect\citeauthoryear{Grimm, Gilfanov, \& Sunyaev}{2003}]{Grimm2003} Grimm H.-J., Gilfanov M., Sunyaev R., 2003, MNRAS, 339, 793 

\bibitem[\protect\citeauthoryear{Grimm, Gilfanov, \& Sunyaev}{2002}]{Grimm2002} Grimm H.-J., Gilfanov M., Sunyaev R., 2002, A\&A, 391, 923 


\bibitem[\protect\citeauthoryear{Goodall, Alouani-Bibi, \& Blundell}{2011}]{Goodall2011} Goodall P.~T., Alouani-Bibi F., Blundell K.~M., 2011, MNRAS, 414, 2838 


\bibitem[\protect\citeauthoryear{Jackson et al.}{2006}]{Jackson2006} Jackson J.~M., et al., 2006, ApJS, 163, 145 

\bibitem[\protect\citeauthoryear{Jiang et al.}{2010}]{Jiang2010} Jiang B., Chen Y., Wang J., Su Y., Zhou X., Safi-Harb S., DeLaney T., 2010, ApJ, 712, 1147 


\bibitem[\protect\citeauthoryear{Kalberla \& Kerp}{2009}]{Kalberla2009} Kalberla P.~M.~W., Kerp J., 2009, ARA\&A, 47, 27 


\bibitem[\protect\citeauthoryear{Kaplan et al.}{2002}]{Kaplan2002} Kaplan D.~L., Kulkarni S.~R., Frail D.~A., van Kerkwijk M.~H., 2002, ApJ, 566, 378 

\bibitem[\protect\citeauthoryear{Khabibullin, Medvedev, \& Sazonov}{2016}]{Khabibullin2016} Khabibullin I., Medvedev P., Sazonov S., 2016, MNRAS, 455, 1414 


\bibitem[\protect\citeauthoryear{Kotani et al.}{1996}]{Kotani1996} Kotani, T., Kawai, N., Matsuoka, M.,\& Brinkmann, W. 1996, PASJ, 48, 619


\bibitem[\protect\citeauthoryear{Koyama et al.}{1996}]{Koyama1996} Koyama K., Maeda Y., Sonobe T., Takeshima T., Tanaka Y., Yamauchi S., 1996, 
PASJ, 48, 249 


\bibitem[\protect\citeauthoryear{Lehmer et al.}{2010}]{Lehmer2010} Lehmer B.~D., Alexander D.~M., Bauer F.~E., Brandt W.~N., Goulding A.~D., Jenkins L.~P., Ptak A., Roberts T.~P., 2010, ApJ, 724, 559 

\bibitem[\protect\citeauthoryear{Liu et al.}{2015}]{Liu2015} 
Liu J.-F., et al., 2015, arXiv, arXiv:1511.09200 



\bibitem[\protect\citeauthoryear{Marshall, Canizares, \& Schulz}{2002}]{Marshall2002} Marshall H.~L., Canizares C.~R., Schulz N.~S., 2002, ApJ, 564, 941

\bibitem[\protect\citeauthoryear{Marshall et al.}{2013}]{Marshall2013} Marshall H.~L., Canizares C.~R., Hillwig T., Mioduszewski A., Rupen M., Schulz N.~S., Nowak M., Heinz S., 2013, ApJ, 775, 75 


\bibitem[\protect\citeauthoryear{McKee \& Ostriker}{2007}]{McKee2007} McKee C.~F., Ostriker E.~C., 2007, ARA\&A, 45, 565 
%
\bibitem[\protect\citeauthoryear{Medvedev 
\& Fabrika}{2010}]{Medvedev2010} Medvedev A., Fabrika S., 2010, MNRAS, 402, 479 

\bibitem[\protect\citeauthoryear{Migliari, Fender, \& M{\'e}ndez}{2002}]{Migliari2002} Migliari S., Fender R., M{\'e}ndez M., 2002, Sci, 297, 1673 


\bibitem[\protect\citeauthoryear{Miller-Jones et al.}{2008}]{Miller2008} Miller-Jones J.~C.~A., Migliari S., Fender R.~P., Thompson T.~W.~J., van der Klis M., M{\'e}ndez M., 2008, ApJ, 682, 1141 


\bibitem[\protect\citeauthoryear{Mineo, Gilfanov, 
\& Sunyaev}{2012}]{Mineo2012} Mineo S., Gilfanov M., Sunyaev R., 2012, MNRAS, 419, 2095 

\bibitem[\protect\citeauthoryear{Molaro, Khatri, \& Sunyaev}{2014}]{Molaro2014} Molaro M., Khatri R., Sunyaev R.~A., 2014, A\&A, 564, A107 


\bibitem[\protect\citeauthoryear{Morrison \& McCammon}{1983}]{Morrison1983} Morrison R., McCammon D., 1983, ApJ, 270, 119 


\bibitem[\protect\citeauthoryear{Ohsuga \& Mineshige}{2014}]{Ohsuga2014} Ohsuga K., Mineshige S., 2014, SSRv, 183, 353 


\bibitem[\protect\citeauthoryear{Panferov \& Fabrika}{1993}]{Panferov1993} Panferov A.~A., Fabrika S.~N., 1993, AstL, 19, 41 


\bibitem[\protect\citeauthoryear{Ponti et al.}{2010}]{Ponti2010} Ponti G., Terrier R., Goldwurm A., Belanger G., Trap G., 2010, ApJ, 714, 732 


\bibitem[\protect\citeauthoryear{Poutanen et al.}{2007}]{Poutanen2007} Poutanen J., Lipunova G., Fabrika S., Butkevich A.~G., Abolmasov P., 2007, MNRAS, 377, 1187 

\bibitem[\protect\citeauthoryear{Rappaport, Podsiadlowski, \& Pfahl}{2005}]{Rappaport2005} Rappaport S.~A., Podsiadlowski P., Pfahl E., 2005, MNRAS, 356, 401 


\bibitem[\protect\citeauthoryear{Revnivtsev et al.}{2004}]{Revnivtsev2004} Revnivtsev M., Sazonov S., Jahoda K., Gilfanov M., 2004, A\&A, 418, 927 

\bibitem[\protect\citeauthoryear{Revnivtsev et al.}{2004}]{RevnivtsevGC2004} Revnivtsev M.~G., et al., 2004, A\&A, 425, L49 

\bibitem[\protect\citeauthoryear{Revnivtsev, Molkov, \& Sazonov}{2006}]{Revnivtsev2006line} Revnivtsev M., Molkov S., Sazonov S., 2006, MNRAS, 373, L11 


\bibitem[\protect\citeauthoryear{Revnivtsev et al.}{2006}]{Revnivtsev2006} Revnivtsev M., Sazonov S., Gilfanov M., Churazov E., Sunyaev R., 2006, A\&A, 452, 169 

\bibitem[\protect\citeauthoryear{Revnivtsev et al.}{2009}]{Revnivtsev2009} Revnivtsev M., Sazonov S., Churazov E., Forman W., Vikhlinin A., Sunyaev R., 2009, Natur, 458, 1142 


\bibitem[\protect\citeauthoryear{Roman-Duval et al.}{2009}]{RD2009} Roman-Duval J., Jackson J.~M., Heyer M., Johnson A., Rathborne J., Shah R., Simon R., 2009, ApJ, 699, 1153 


\bibitem[\protect\citeauthoryear{Roman-Duval et al.}{2010}]{RD2010} Roman-Duval J., Jackson J.~M., Heyer M., Rathborne J., Simon R., 2010, ApJ, 723, 492 


\bibitem[\protect\citeauthoryear{Sabin et al.}{2013}]{Sabin2013} Sabin L., et al., 2013, MNRAS, 431, 279 

\bibitem[\protect\citeauthoryear{Sazonov et al.}{2008}]{Sazonov2008} Sazonov S., Krivonos R., Revnivtsev M., Churazov E., Sunyaev R., 2008, A\&A, 482, 517 

\bibitem[\protect\citeauthoryear{Sazonov, Lutovinov, \& Krivonos}{2014}]{Sazonov2014} Sazonov S.~Y., Lutovinov A.~A., Krivonos R.~A., 2014, AstL, 40, 65 


\bibitem[\protect\citeauthoryear{Shakura \& Sunyaev}{1973}]{SS1973} Shakura N.~I., Sunyaev R.~A., 1973, A\&A, 24, 337

\bibitem[\protect\citeauthoryear{Su et al.}{2011}]{Su2011} Su Y., Chen Y., Yang J., Koo B.-C., Zhou X., Lu D.-R., Jeong I.-G., DeLaney T., 2011, ApJ, 727, 43 

\bibitem[\protect\citeauthoryear{Sugizaki et al.}{2001}]{Sugizaki2001} Sugizaki M., Mitsuda K., Kaneda H., Matsuzaki K., Yamauchi S., Koyama K., 2001, ApJS, 134, 77 

\bibitem[\protect\citeauthoryear{Sun et al.}{2011}]{Sun2011} Sun X.~H., Reich P., Reich W., Xiao L., Gao X.~Y., Han J.~L., 2011, A\&A, 536, A83 


\bibitem[\protect\citeauthoryear{Sunyaev, Markevitch, \& Pavlinsky}{1993}]{Sunyaev1993} Sunyaev R.~A., Markevitch M., Pavlinsky M., 1993, ApJ, 407, 606 

\bibitem[\protect\citeauthoryear{Sunyaev \& Churazov}{1996}]{Sunyaev1996} Sunyaev R.~A., Churazov E.~M., 1996, AstL, 22, 648 


\bibitem[\protect\citeauthoryear{Sunyaev \& Churazov}{1998}]{Sunyaev1998} Sunyaev R., Churazov E., 1998, MNRAS, 297, 1279 

\bibitem[\protect\citeauthoryear{Sunyaev, Uskov, \& Churazov}{1999}]{Sunyaev1999} Sunyaev R.~A., Uskov D.~B., Churazov E.~M., 1999, AstL, 25, 199 

\bibitem[\protect\citeauthoryear{Urquhart 
\& Soria}{2015}]{Urquhart2015} Urquhart R., Soria R., 2015, arXiv, arXiv:1511.05275 

\bibitem[\protect\citeauthoryear{Vallee}{1995}]{Vallee1995} Vallee J.~P., 1995, ApJ, 454, 119 

\bibitem[\protect\citeauthoryear{Vinokurov, Fabrika, \& Atapin}{2013}]{Vinokurov2013} Vinokurov A., Fabrika S., Atapin K., 2013, Astron. Bull., 68, 139 


\bibitem[\protect\citeauthoryear{Walton et al.}{2014}]{Walton2014} Walton D.~J., et al., 2014, ApJ, 793, 21 

\bibitem[\protect\citeauthoryear{Worrall et al.}{1982}]{Worrall1982} Worrall D.~M., Marshall F.~E., Boldt E.~A., Swank J.~H., 1982, ApJ, 255, 111 

\bibitem[\protect\citeauthoryear{Yamamoto et al.}{2008}]{Yamamoto2008} Yamamoto H., et al., 2008, PASJ, 60, 715 


\end{thebibliography}
\end{document}